\documentclass[12pt,journal,onecolumn]{IEEEtran}

\usepackage[ansinew]{inputenc}
\usepackage[dvips]{graphicx}
\usepackage[T1]{fontenc}
\usepackage{amsmath,amssymb}

\DeclareMathOperator{\Br}{Br}
\DeclareMathOperator{\Rad}{Rad}
\DeclareMathOperator{\Gal}{Gal}

\newtheorem{thm}{Theorem}[section]
\newtheorem{proposition}[thm]{Proposition}
\newtheorem{lemma}[thm]{Lemma}
\newtheorem{corollary}[thm]{Corollary}

\newtheorem{definition}{Definition}[section]
\newtheorem{exam}{Example}[section]
\newtheorem{remark}{Remark}[section]

\newcommand{\Z}{{\mathbf{Z}}}
\newcommand{\Q}{{\mathbf{Q}}}

\newcommand{\C}{{\mathbf{C}}}

\newcommand{\A}{{\mathcal{A}}}
\newcommand{\B}{{\mathcal{B}}}
\newcommand{\D}{{\mathcal{D}}}

\newcommand{\G}{{\mathcal{G}}}
\newcommand{\I}{{\mathcal{I}}}
\newcommand{\J}{{\mathcal{J}}}
\newcommand{\OO}{{\mathcal{O}}}
\newcommand{\PP}{{\mathcal{P}}}

\begin{document}

\title{On the Densest MIMO Lattices from Cyclic Division Algebras
\thanks{This work was supported in part by the Nokia Foundation, the
Foundation of Technical Development, Finland, the Foundation of the Rolf
Nevanlinna Institute, Finland, and the Academy of Finland, grant \#108238.}}

\author{Camilla Hollanti, \thanks{C. Hollanti and R. Vehkalahti are with the 
Laboratory of Discrete Mathematics for Information Technology, Turku Centre
for Computer Science, Joukahaisenkatu 3-5 B, FIN-20520 Turku, Finland (e-mail:
\{cajoho, roiive\}@utu.fi).}
Jyrki Lahtonen, \textit{Member, IEEE}, \thanks{J. Lahtonen and K. Ranto are
with the Department of Mathematics, FIN-20014 University of Turku, Finland
(e-mail: \{lahtonen, kara\}@utu.fi).} Kalle Ranto, and Roope Vehkalahti}

\maketitle

\begin{abstract}
It is shown why the discriminant of a maximal order within a cyclic division
algebra must be minimized in order to get the densest possible matrix
lattices with a prescribed nonvanishing minimum determinant. Using results
from class field theory a lower bound to the minimum discriminant of a maximal
order with a given center and index (= the number of Tx/Rx antennas) is
derived. Also numerous examples of division algebras achieving our bound are
given. E.g. we construct a matrix lattice with QAM coefficients that has
2.5 times as many codewords as the celebrated Golden code of the same minimum
determinant. We describe a general algorithm due to Ivanyos and R\'onyai for
finding maximal orders within a cyclic division algebra and discuss our
enhancements to this algorithm. We also consider general methods for finding
cyclic division algebras of a prescribed index achieving our lower bound.
\end{abstract}


\section{Overview}
\label{overview}

Multiple-antenna wireless communication promises very high data rates, in
particular when we have perfect channel state information (CSI) available at
the receiver. In \cite{GFBK} the design criteria for such systems were
developed, and further on the evolution of space-time (ST) codes took  two
directions: trellis codes and block codes. Our work concentrates on the latter
branch.

We are interested in the coherent multiple input-multiple output
(MIMO) case. A {\it lattice} is a discrete finitely generated free abelian
subgroup $\mathbf{L}$ of a real or complex finite dimensional
vector space $\bf{V}$, called the ambient space. In the space-time
setting a natural ambient space is the space ${\cal M}_n(\C)$ of
complex $n\times n$ matrices. We only consider full rank lattices
that have a basis $x_1,x_2,\ldots, x_{2n^2}$ consisting of matrices
that are linearly independent over the field of real numbers. We can
form a $2n^2\times 2n^2$ matrix $M$ having rows consisting of the
real and imaginary parts of all the basis elements. It is well known
that the measure, or hypervolume, $m(\mathbf{L})$ of the fundamental
parallelotope of the lattice then equals the absolute value of $\det(M)$.
Alternatively we may use the {\it Gram matrix}
\[G(\mathbf{L})=MM^T=\left(\Re tr(x_ix_j^H)\right)_{1\le i,j\le 2n^2},\]
where $H$ indicates the complex conjugate transpose of a matrix. The Gram
matrix then has a positive determinant equal to $m(\mathbf{L})^2.$

From the pairwise error probability (PEP) point of view \cite{TSC}, the
performance of a space-time code is dependent on two parameters: {\it
diversity gain} and {\it coding gain}. Diversity gain is the minimum of the
rank of the difference matrix $X-X'$ taken over all distinct code matrices
$X,X'\in\mathcal{C}$, also called the {\it rank} of the code $\mathcal{C}$.
When $\mathcal{C}$ is full-rank, the coding gain is proportional to the
determinant of the matrix $(X-X')(X-X')^H$. The minimum of this determinant
taken over all distinct code matrices is called the {\it minimum determinant}
of the code $\mathcal{C}$. If it is bounded away from zero even in the limit 
as SNR $\rightarrow\infty$, the ST code is said to have the {\it nonvanishing
determinant} (NVD) property \cite{BRV}. For non-zero square matrices, being
full-rank coincides with being invertible.

The  {\it data rate} $R$  in symbols per channel use  is given by
\[R=\frac{1}{n}\log_{|S|}(|\mathcal{C}|),\]
where $|S|$ and $|\mathcal{C}|$ are the sizes of the symbol set and code
respectively. This is not to be confused with the {\it rate of a code design}
defined as the ratio of the number of transmitted information symbols to the
decoding delay (equivalently, block length) of these symbols at the receiver
for any given number of transmit antennas  using any complex signal
constellations. If this ratio is equal to the delay, the code is said to have
{\it full rate}.

The very first STBC for two transmit antennas was the  {\it Alamouti code}
\cite{Alam} representing multiplication in the ring of quaternions. As the
quaternions form a division algebra, such matrices must be invertible, i.e.
the resulting STBC meets the rank criterion. Matrix representations of other
division algebras have been proposed as STBCs at least in
\cite{HL2}-\cite{EKPKL}, and (though without explicitly saying so) \cite{WX}.
The most recent work \cite{SRS}-\cite{WX} has concentrated on adding
multiplexing gain, i.e. multiple input-multiple output (MIMO)  applications,
and/or combining it with a good minimum determinant. It has been shown in
\cite{EKPKL} that CDA-based square ST codes with the NVD property achieve the
diversity-multiplexing gain (D-MG) tradeoff introduced in \cite{ZT}. The codes
proposed in this paper all fall into this category  and are in that sense
optimal. Furthermore, algebras with an imaginary quadratic field as a center
yield lattices with a good minimum determinant, as the corresponding rings of
integers have no short non-zero elements.

Here, yet another design criterion is brought  into the playground, namely an
explicit criterion for maximizing the density of the code. The field of ST
coding seems to be lacking a general, precise notion for the density in the
case of noncommutative structures. In fact, according to our best knowledge
the theory of orders required for giving this notion has never been considered
before in this area.

Hence, after a cyclic division algebra has been chosen, the next step is to
choose a corresponding lattice, or what amounts to the same thing, to choose
an order within the algebra. Most authors \cite{WX}, \cite{EKPKL} have gone
with the so-called natural order (see the next section for a definition). One
of the points we want to emphasize in this article is to use the maximal
orders instead. The idea is that one can sometimes use several cosets of the
natural order without sacrificing anything in terms of the minimum
determinant. So the study of maximal orders is clearly motivated by an analogy
from the theory of error correcting codes: why one would use a particular
code of a given minimum distance and length, if a larger code with the same
parameters is available. The standard matrix representation of the natural
order results in codes that have a so-called threaded layered structure
\cite{GH}. When a maximal order is used, the code will then also extend
`between layers'. However, our simulations suggest that restoring the layered
structure somewhat by replacing the maximal order with its smartly chosen
ideal yields codes with better performance. For more details about this see
Section \ref{simulaatiosektio} below. Earlier we have successfully used
maximal orders in a construction of some 4Tx antenna MISO lattices \cite{HL2}.

In some cases the index of the natural order as a sublattice of a maximal
order is quite large. E.g. in the cases of a family of cyclic algebras
suggested in \cite{KR} one can theoretically increase the data rate by $1.5$,
$6.5$ and $20.5$ bits per channel use for $2$, $4$ and $8$ antenna codes,
respectively. We do emphasize that such increments of data rates are only
theoretical in nature. This is because one is compelled to use relatively
large subsets of the infinite lattice before the full density advantage of the
maximal order is attained. Also the lattice of a fully multiplexing 8Tx+8Rx
antenna MIMO code has dimension $128$. The nearest vector problem in such
high-dimensional lattices is used in some cryptographic applications, so it is
safe to say that ML-decoding of such lattices will have prohibitive
complexity. These numbers, however, motivated us to look for methods of
locating maximal orders. A general purpose algorithm for this task has been
developed by Ivanyos and R\'onyai \cite{IR}. A commercially available version
of their algorithm is implemented by W. van de Graaf as part of the computer
algebra system MAGMA \cite{Mag}. It turned out that this general purpose
algorithm was not able to handle the algebras of index eight. To deal with
these special cases we developed some enhancements to their algorithm.

Given that maximal orders provide the best codes in terms of minimum
determinant vs. average power we are left with the  question: Which division
algebra should we use? To continue the analogy from the theory of
error-correcting codes we want to find the codes with the highest possible
density. That is, with the smallest fundamental parallelotope. To that end we
need a suitable tool for parameterizing the cyclic division algebras with a
given center and index. Luckily, relatively deep results from class field
theory provide us with the necessary tool of Hasse invariants. The measure of
a fundamental parallelotope of a maximal order (that will later on be
referred to as the discriminant of the division algebra) can be expressed in
terms of Hasse invariants \cite{R}. With these results at hand we then derive
a lower bound to the discriminant. While the proof of the lower bound is not
constructive per se, it does show that our lower bound is achievable. In the
latter parts of this article we describe some techniques for constructing
division algebras with a minimal discriminant.

It is worth mentioning that in \cite{FOV} the authors have made a similar
approach in the reduced case of commutative number fields.

While our interest in these problems is mostly theoretical, some of the
densest lattices we have found also perform well in computer simulations. Our
construction of the densest $2\times 2$ matrix lattice improves upon the
deservedly celebrated Golden code in block error rates by about $0.9$ dB at
data rates from $5$ to $6$ bpcu. The performance of both the rival codes can
be further improved by coset optimization and this also cuts down the gap to
about $0.3$ dB. Observe that at the data rate of 4 bpcu we have a tie. This is
easily explained by the fact that for codes of that size there is a
particularly attractive choice for the coset of the Golden code. Another point
worth keeping in mind is that the somewhat irregular geometry of our lattice
more or less necessitates the use of a code book as opposed to a simple
combination of Gray coding and PAM. However, this also holds for the Golden
code, when we do any coset optimization. Thus we might conclude that our work
shows that not using a codebook costs about $1$ dB.

The paper is organized as follows. In Section \ref{CA}, various algebraic
notions related to cyclic algebras, Brauer groups, orders, discriminants, and
localizations are introduced and demonstrated by examples. Furthermore, it is
shown that maximizing the density of the code, i.e. minimizing the fundamental
parallelotope is equivalent to minimizing the discriminant. This leads us to
Section \ref{DB}, where we derive an achievable lower bound for the
discriminant. In Section \ref{Finding}, we propose a general algorithm due to
Ivanyos and R\'onyai \cite{IR} for finding maximal orders. Unfortunately, when
we were trying to use the MAGMA implementation of this algorithm for finding
maximal orders of certain cyclic division algebra of index no more than $8$,
the memory of a typical modern PC turned out to be insufficient. Hence, also
some enhancements to their algorithm in this special case are discussed here.
The Perfect codes are analyzed in Section \ref{Perfect} in terms of Hasse
invariants and discriminants. We show that the natural orders (i.e. the orders
the authors have used in \cite{BORV}) of the related algebras are maximal in
the cases of $\#Tx=2$ and $\#Tx=3$, but can be enlarged in the cases of
$\#Tx=4$ and $\#Tx=6$. In Section \ref{Constructing} we    construct division
algebras with a minimal discriminant. The case of a unit non-norm element is
separated from the general construction. Finally in Section
\ref{simulaatiosektio}, the theory is brought into practice by giving an
explicit code construction that outperforms or ties with the Golden code.
Simulation results are provided to back up this claim.

\section{Cyclic algebras, Brauer groups, orders, and discriminants}
\label{CA}

We refer the interested reader to \cite{Jac} and \cite{SRS} for a detailed
exposition of the theory of simple algebras, cyclic algebras, their matrix
representations and their use in ST-coding. We only recall the basic
definitions and notations here. In the following, we consider number field
extensions $E/F$, where $F$ denotes the base field and $F^*$ (resp. $E^*$)
denotes the set of the non-zero elements of $F$ (resp. $E$). In the
interesting cases $F$ is an imaginary quadratic field, either $\Q(i)$ or
$\Q(\sqrt{-3})$. We assume that $E/F$ is a cyclic field extension of degree
$n$ with Galois group $\Gal(E/F)=\left\langle \sigma\right\rangle$. Let
$\A=(E/F,\sigma,\gamma)$ be the corresponding cyclic algebra of degree $n$
($n$ is also called the {\it index} of $\A$), that is
\begin{center}
$\A=E\oplus uE\oplus u^2E\oplus\cdots\oplus u^{n-1}E$,
\end{center}
as a (right) vector space over $E$. Here $u\in\A$ is an auxiliary generating
element subject to the relations $xu=u\sigma(x)$ for all $x\in E$ and
$u^n=\gamma\in F^*$. An element
$a=x_0+ux_1+\cdots+u^{n-1}x_{n-1}\in\mathcal{A}$ has the following
representation as a matrix $A=$
\[\begin{pmatrix}
x_0& \gamma\sigma(x_{n-1})& \gamma\sigma^2(x_{n-2})&\cdots &
\gamma\sigma^{n-1}(x_1)\\
x_1&\sigma(x_0)&\gamma\sigma^2(x_{n-1})& &\gamma\sigma^{n-1}(x_2)\\
x_2& \sigma(x_1)&\sigma^2(x_0)& &\gamma\sigma^{n-1}(x_3)\\
\vdots& & & & \vdots\\
x_{n-1}& \sigma(x_{n-2})&\sigma^2(x_{n-3})&\cdots&\sigma^{n-1}(x_0)\\
\end{pmatrix}.\]
We refer to this as the standard matrix representation of $\A$. Observe that
some variations are possible here. E.g.~one may move the coefficients $\gamma$
from the upper triangle to the lower triangle by conjugating this matrix with
a suitable diagonal matrix. Similarly one may arrange to have the first row to
contain the ``pure'' coefficients $x_0,\ldots,x_{n-1}$. Such changes do not
affect the minimum determinant nor the density of the resulting lattices.

If we denote the basis of $E$ over  $\OO_F$ by $\{1,e_1,...,e_{n-1}\}$, then 
the elements $x_i,\ i=0,...,n-1$ in the above matrix take the form
$x_i=\sum_{k=0}^{n-1}f_ke_k$, where $f_k\in \OO_F$ for all $k=0,...,n-1$.
Hence $n$ complex symbols are transmitted per channel use, i.e. the design has
rate $n$. In literature this is often referred to as having a {\it full rate}.

\begin{definition}
The determinant (resp. trace) of the matrix $A$ above is called the {\it
reduced norm} (resp. {\it reduced trace}) of the element $a\in\A$ and is
denoted by $nr(a)$ (resp. $tr(a)$).
\end{definition}

\begin{remark}
\label{red vs usual} 
The connection with the usual norm map $N_{\A/F}(a)$ (resp. trace map
$T_{\A/F}(a)$) and the reduced norm $nr(a)$ (resp. reduced trace $tr(a)$) of
an element $a\in\A$ is $N_{\A/F}(a)=(nr(a))^n$ (resp. $T_{\A/F}(a)=n tr(a)$),
where $n$ is the degree of $E/F$.
\end{remark}

\begin{definition}
\label{central simple}
An algebra $\A$ is called {\it simple} if it has no nontrivial ideals. An
$F$-algebra $\A$ is {\it central} if its center $Z(\A)=\{a\in\A\ |\ aa'=a'a\
\forall a'\in\A\}=F$.
\end{definition}

\begin{definition}
\label{radical} 
Let $S$ denote an arbitrary ring with identity. The {\it Jacobson radical} of
the ring $S$ is the set $\Rad(S)=$
\[\{x\in S\ |\ xM=0 \ \textrm{for all simple left } S\textrm{-modules}\ M\}.\]
\end{definition}

$\Rad(S)$ is a two-sided ideal in $S$ containing every nilpotent (i.e. for
which $\I^k=0$ for some $k\in \Z_+$) one-sided ideal $\I$ of $S$. Also,
$\Rad(S)$ can be characterized as the intersection of the maximal left ideals
in $S$. If $S$ is a finite dimensional algebra over a field or, more
generally, left or right Artinian then $\Rad(S)$ is the maximal nilpotent
ideal in $S$.

A division algebra may be represented as a cyclic algebra in many ways as
demonstrated by the following example.

\begin{exam}
\label{goldenalgebra}
The division algebra $\mathcal{GA}$ used in \cite{BRV} to construct the
Golden code is gotten as a cyclic algebra with $F=\Q(i)$, $E=\Q(i,\sqrt{5})$,
$\gamma=i$, when the $F$-automorphism $\sigma$ is determined by
$\sigma(\sqrt{5})=-\sqrt{5}$. We also note that in addition to this
representation $\mathcal{GA}$ can be given another construction as a cyclic
algebra. As now $u^2=i$ we immediately see that $F(u)$ is a subfield of
$\mathcal{GA}$ that is isomorphic to the eighth cyclotomic field
$E'=\Q(\zeta)$, where $\zeta=(1+i)/\sqrt{2}$. The relation
$u\sqrt{5}=-\sqrt{5}u$ read differently means that we can view $u$ as the
complex number $\zeta$ and $\sqrt{5}$ as the auxiliary generator, call it
$u'=\sqrt{5}$. We thus see that the cyclic algebra
\[E'\oplus u'E'=(E'/F,\sigma',\gamma')\]
is isomorphic to the Golden algebra. Here $\sigma'$ is the $F$-automorphism
of $E'$ determined by $\zeta\mapsto -\zeta$ and $\gamma'=u'^2=5$.
\end{exam}

Any cyclic algebra is a central simple $F$-algebra (cf. Definition
\ref{central simple}). Two central simple $F$-algebras $\A$ and $\B$
are said to be {\it similar}, if there exist integers $m$ an $n$ such that the
matrix algebras $\mathcal{M}_n(\A)$ and $\mathcal{M}_m(\mathcal{B})$ are
isomorphic $F$-algebras. Wedderburn's structure theorem \cite[Theorem, p.
171]{Jac} tells us that any central simple algebra is a matrix algebra over a
central simple division algebra, and it easily follows that within any
similarity class there is a unique division algebra. Similarity classes of
central simple algebras form a group (under tensor product over $F$), called
the {\it Brauer group} $\Br(F)$ of the field $F$. If $F'$ is an extension
field of $F$, and $\A$ is a central simple $F$-algebra, then the tensor
product $\A'=\A\otimes_FF'$ is a central simple $F'$-algebra. We refer to this
algebra as the algebra gotten from $\A$ by {\it extending the scalars to
$F'$}.

The next proposition due to A.~A.~Albert \cite[Theorem 11.12, p. 184]{AA}
tells us when a cyclic algebra is a division algebra.

\begin{proposition}[Norm condition]
\label{albert}
The cyclic algebra $\A=(E/F,\sigma,\gamma)$ of degree $n$ is a division
algebra if and only if the smallest factor $t\in\Z_+$ of $n$ such that
$\gamma^t$ is the norm of some element of $E^*$ is $n$.
\end{proposition}

Due to the above proposition, the element $\gamma$ is often referred to as
the {\it non-norm element}.

Let $F$ be an algebraic number field that is finite dimensional over $\Q$.
Denote its ring of integers by $\OO_F$. If $P$ is a prime ideal of $\OO_F$, we
denote the $P$-adic completion of $F$ by $\hat F_P$. The division algebras
over $\hat F_P$ are easy to describe. They are all gotten as cyclic algebras
of the form $\A(n,r)=(E/\hat F_P,\sigma,\pi^r)$, where $E$ is the unique
unramified extension of $\hat F_P$ of degree $n$, $\sigma$ is the Frobenius
automorphism, and $\pi$ is the prime element of $F_P$. The quantity $r/n$ is
called the {\it Hasse invariant} of this algebra and $n$ is referred to as
the {\it local index}. It immediately follows from Proposition \ref{albert}
that $\A(n,r)$ is a division algebra, if and only if $(r,n)=1$. For a
description of the theory of Hasse invariants we refer the reader to
\cite[p. 266]{R} or \cite{M}.

We are now ready to present some of the basic definitions and results from
the theory of maximal orders. The general theory of maximal orders can be
found in \cite{R}.

Let $R$ denote a Noetherian integral domain with a quotient field $F$, and let
$\A$ be a finite dimensional $F$-algebra.

\begin{definition}
An $R$-{\it order} in the $F$-algebra $\A$ is a subring $\Lambda$ of $\A$,
having the same identity element as $\A$, and such that $\Lambda$ is a
finitely generated module over $R$ and generates $\A$ as a linear space over
$F$. An order $\Lambda$ is called {\it maximal}, if it is not properly
contained in any other $R$-order.
\end{definition}

Let us illustrate the above definition by concrete examples.

\begin{exam}
\label{left order}
(a) Orders always exist: If $M$ is a {\it full} $R$-lattice in $\A$, i.e.
$FM=\A$, then the {\it left order} of $M$ defined as $\OO_l(M)=\{x\in\A\ |\
xM\subseteq M\}$ is an $R$-order in $\A$. The right order is defined in an
analogous way.

(b) If $R$ is the ring of integers $\OO_F$ of the number field $F$, then the
ring of integers $\OO_E$ of the extension field $E$ is the unique maximal
order in $E$. For example, in the case of the cyclotomic field $E=\Q(\zeta)$,
where $\zeta=\exp(2\pi i/k)$ is a primitive root of order $k$ the maximal
order is $\OO_E=\Z[\zeta]$.
\end{exam}

The next proposition (see \cite[proof of Theorem 3.2]{Ron}) is useful when
computing left orders in Section \ref{Finding}.

\begin{proposition}
\label{ronyaialg 3.2} 
Let $\A$ be a simple algebra over $F$ and  $M$ a finitely generated
$\OO_F$-module such that $FM=\A$. Then there exists an element
$s\in\OO_F\setminus\{0\}$ such that $s\cdot 1\in M$. Moreover,
$\OO_l(M)=\{b\in s^{-1}M\ |\ bM\leq M\}\leq s^{-1}M$.
\end{proposition}

For the purposes of constructing MIMO lattices the reason for concentrating
on orders is summarized in the following proposition (e.g. \cite[Theorem 10.1,
p. 125]{R}). We simply rephrase it here in the language of MIMO-lattices. We
often (admittedly somewhat inaccurately) identify an order (or its subsets)
with its standard matrix representation.

\begin{proposition}
Let $\Lambda$ be an order in a cyclic division algebra $(E/F,\sigma,\gamma)$.
Then for any non-zero element $a\in\Lambda$ its reduced norm $nr(a)$ is a
non-zero element of the ring of integers $\OO_F$ of the center $F$. In
particular, if $F$ is an imaginary quadratic number field, then the minimum
determinant of the lattice $\Lambda$ is equal to one.
\end{proposition}

\begin{exam}
\label{naturalorder} 
In any cyclic algebra we can always choose the element $\gamma\in F^*$ to
be an algebraic integer. We immediately see that the $\OO_F$-module
\[\Lambda=\OO_E\oplus u \OO_E\oplus\cdots\oplus u^{n-1}\OO_E,\]
where $\OO_E$ is the ring of integers, is an $\OO_F$-order in the cyclic
algebra $(E/F,\sigma,\gamma)$. We refer to this $\OO_F$-order as the {\it
natural order}. It will also serve as a starting point when searching for
maximal orders.

We want the reader to note that in any central simple algebra a maximal
$\Z$-order is a maximal $\OO_F$-order as well. Note also that if $\gamma$ is
not an algebraic integer, then $\Lambda$ fails to be closed under
multiplication. This may adversely affect the minimum determinant of the
resulting matrix lattice, as elements not belonging to an order may have
non-integral (and hence small) norms.
\end{exam}

We remark that the term `natural order' is somewhat misleading. While it is
the first order that comes to mind, there is nothing canonical about it.
Indeed, distinct realizations of a given division algebra as a cyclic algebra
often lead to different natural orders. E.g. constructing the algebra of
rational Hamiltonian quaternions from the cyclic extension $\Q(\sqrt{-3})/\Q$
as opposed to the more common $\Q(i)/\Q$ leads to a different natural order.
The interested reader may verify this as an exercise by starting with the
observation that the Hamiltonian quaternion $i+j+k$ may be used as a square
root of $-3$.

\begin{definition}
\label{discriminant} 
Let $m=dim_F\A$. The {\it discriminant} of the $R$-order $\Lambda$ is the
ideal $d(\Lambda/R)$ in $R$ generated by the set
\[\{\det(tr(x_ix_j))_{i,j=1}^m\ |\ (x_1,...,x_m)\in\Lambda^m\}.\]
\end{definition}

In the interesting cases of $F=\Q(i)$ (resp. $F=\Q(\sqrt{-3})$) the ring
$R=\Z[i]$ (resp. $R=\Z[\omega]$, $\omega=(-1+\sqrt{-3})/2$) is a Euclidean
domain, so in these cases (as well as in the case $R=\Z$) it makes sense to
speak of the discriminant as an element of $R$ rather than as an ideal. We
simply pick a generator of the discriminant ideal, and call it the
discriminant. Equivalently we can compute the discriminant as
\[d(\Lambda/R)= \det(tr(x_ix_j))_{i,j=1}^m,\]
where $\{x_1,\ldots,x_m\}$ is any $R$-basis of $\Lambda$. It is readily
seen that whenever $\Lambda\subseteq\Gamma$ are two $R$-orders, then
$d(\Gamma)$ is a factor of $d(\Lambda)$. The index $[\Gamma:\Lambda]$  is
related to discriminants by the following lemma.

\begin{lemma}
\label{diskriminanttiindeksi}
\[[R:d(\Lambda)R]=[\Gamma : \Lambda]^2 [R:d(\Gamma)R]\]
\end{lemma}

\begin{proof}
\cite[p.66]{R}
\end{proof}

It turns out (cf. \cite[Theorem 25.3]{R}) that all the maximal orders of a
division algebra share the same discriminant that we will refer to as the
discriminant of the division algebra. In this sense a maximal order has the
smallest possible discriminant among all orders within a given division
algebra, as all the orders are contained in the maximal one.

For an easy  reference we also note the following basic formula for the
discriminant of certain cyclotomic fields.

\begin{proposition}
\label{syklotomidiskriminantti}
Let $\zeta_\ell=\exp(2\pi i/2^\ell)$ be a complex primitive root of unity of
order $2^\ell$, where $\ell\ge2$ is an integer. Then
$n=[\Q(\zeta_\ell):\Q(i)]=2^{\ell-2}$ and
\[d(\Z[\zeta_\ell]/\Z[i])=(1+i)^{2n(\ell-2)}.\]
\end{proposition}

The definition of the discriminant closely resembles that of the Gram matrix
of a lattice, so the following results are unsurprising and probably well
known. We include them for lack of a suitable reference.

\begin{lemma}
Assume that $F$ is an imaginary quadratic number field and that $1$ and
$\theta$ form a $\Z$-basis of its ring of integers $R$. Assume further that
the order $\Lambda$ is a free $R$-module (an assumption automatically
satisfied, when $R$ is a principal ideal domain). Then the measure of the
fundamental parallelotope equals
\[m(\Lambda)=\vert \Im\theta \vert^{n^2} \vert d(\Lambda /R)\vert.\]
\end{lemma}

\begin{proof}
Let $A=(a_{ij})$ be an $n\times n$ complex matrix. We flatten it out into
a $2\times 2n^2$ matrix $L(A)$ by first forming a vector of length $n^2$ out
of the entries (e.g. row by row) and then replacing a complex number $z$
by a diagonal two by two matrix with entries $z$ and $z^*$ (= the usual
complex conjugate of $z$). If $A$ and $B$ are two square matrices with $n$
rows we can easily verify the  identities
\begin{equation}\label{eq:L(A)H}
L(A) L(B)^H=\begin{pmatrix}tr(AB^H)&0\\0& tr(A^HB)\end{pmatrix}
\end{equation}
and
\begin{equation}\label{eq:L(A)T}
L(A) L(B^T)^T=\begin{pmatrix}tr(AB)&0\\0& tr(AB)^*\end{pmatrix}.
\end{equation}

Next let $\B=\{x_1,x_2,\ldots,x_{n^2}\}$ be an $R$-basis for $\Lambda$.
We form the $2n^2\times 2n^2$ matrix $L({\mathcal B})$ by stacking the
matrices $L(x_i)$ on top of each other. Similarly we get $R(\B)$ by
using the matrices $L(x_i^T)^T$ as `column blocks'. Then by \eqref{eq:L(A)T}
the matrix $M=L(\B)R(\B)$ consists of two by two blocks of the form
\[L(x_i)L(x_j^T)^T=\begin{pmatrix}tr(x_ix_j)&0\\0&
tr(x_ix_j)^*\end{pmatrix}.\]
Clearly $\det R(\B) = \pm\det L(\B)$, and $\det M=\vert d(\Lambda/R)\vert^2$,
so we get 
\[\vert d(\Lambda/R)\vert = \vert \det L(\B)\vert.\]

Next we turn our attention to the Gram matrix. By our assumptions the set
${\mathcal B} \cup \theta {\mathcal B}$ is a $\Z$-basis for $\Lambda$.
Let us denote
\[D=\begin{pmatrix}1&1\\ \theta&\theta^* \end{pmatrix}.\]
From the identities $\Re (xy^*)=(xy^*+x^*y)/2$ and
\[D\begin{pmatrix}x&0\\ 0&x^* \end{pmatrix}
=\begin{pmatrix}x&x^*\\ \theta x&\theta^* x^* \end{pmatrix}\]
together with \eqref{eq:L(A)H} it follows that for
any two $n\times n$ matrices $A$ and $B$ we have
\[\frac{1}{2}\left(D L(A)\right)\left(D L(B)\right)^H
=\begin{pmatrix}
\Re(tr(AB^H))&\Re(tr(A(\theta B)^H)\\
\Re(tr(\theta A B^H))&\Re(tr(\theta A (\theta B)^H))
\end{pmatrix}.\]

Therefore, if we denote by $D^{[n]}$ the $2n^2\times 2n^2$ matrix having $n^2$
copies of $D$ along the diagonal and zeros elsewhere, we get the following
formula for the Gram matrix
\[G(\Lambda)=\frac{1}{2}\left(D^{[n]} L({\mathcal B})\right)\left(D^{[n]}
L(\B)\right)^H.\]
Thus,
\[m(\Lambda)=\det G(\Lambda)^{1/2}=\left\vert\det L(\B)\right\vert
\left\vert\frac{1}{2}\det D\right\vert^{n^2}.\]
Our claim now follows from all these computations and the fact that
$(\det D)/2=(\theta^*-\theta)/2=-\Im \theta$.
\end{proof}

In the respective cases $F=\Q(i)$ and $F=\Q(\sqrt{-3})$ we have $\theta=i$ and
$\theta=(-1+\sqrt{-3})/2$ respectively, so we immediately get the following
two corollaries.

\begin{corollary}
\label{gausskoppi}
Let $F=\Q(i), R=\Z[i]$, and assume that $\Lambda\subset (E/F,\sigma,\gamma)$
is an $R$-order. Then the measure of the fundamental parallelotope equals
\[m(\Lambda)=\vert d(\Lambda/\Z[i])\vert.\]
\end{corollary}

\begin{exam}
When we scale the Golden code \cite{BRV} to have a unit minimum determinant,
all the $8$ elements of its $\Z$-basis will have length $5^{1/4}$ and the
measure of the fundamental parallelotope is thus $25$. In view of all of the
above this is also a consequence of the fact that the $\Z[i]$-discriminant of
the natural order of the Golden algebra is equal to $25$. As was observed in
\cite{HL} the natural order happens to be maximal in this case, so the Golden
code cannot be improved upon by enlarging the order within $\mathcal{GA}$.
\end{exam}

\begin{corollary}
\label{eisensteinkoppi}
Let $\omega=(-1+\sqrt{-3})/2$, $F=\Q(\omega)$, $R=\Z[\omega]$, and assume
that $\Lambda\subset (E/F,\sigma,\gamma)$ is an $R$-order. Then the measure of
the fundamental parallelotope equals
\[m(\Lambda)=(\sqrt{3}/2)^{n^2}\vert d(\Lambda/\Z[\omega])\vert.\]
\end{corollary}

The upshot is that in both cases \textbf{maximizing the density of the code,
i.e. minimizing the fundamental parallelotope, is equivalent to minimizing the
discriminant}. Thus, in order to get the densest MIMO-codes we need to look
for division algebras that have a maximal order with as small a discriminant
as possible.

For an easy reference we also include the following result that is a
relatively easy consequence of the definitions.

\begin{lemma}
\label{naturaldiscriminant}
Let $E/F$ be as above, assume that $\gamma$ is an algebraic integer of $F$,
and let $\Lambda$ be the natural order of Example \ref{naturalorder}. If
$d(E/F)$ is the $\OO_F$-discriminant of $\OO_E$ (often referred to as the
relative discriminant of the extension $E/F$), then
\[d(\Lambda/\OO_F)=d(E/F)^n \gamma^{n(n-1)}.\]
\end{lemma}

\begin{proof}
In the expansion
\[\Lambda=\OO_E\oplus u \OO_E\oplus\cdots\oplus u^{n-1}\OO_E\]
we see that $u^i\OO_E$ and $u^j\OO_E$ are orthogonal to each other with
respect to the bilinear form given by the reduced trace except in the cases
where $i+j\equiv 0 \pmod{n}$. Assume that $i+j$ is divisible by $n$ for some
$i,j$ in the range $0\le i,j<n$, and that $x_1,\ldots,x_n$ are elements of
${\cal O}_E$. Then the multiplication rules of the cyclic algebra imply that
\[\det(tr(u^ix_k u^jx_\ell))_{k,\ell=1}^n =\pm\det(u^{i+j} tr(x_k
x_\ell))_{k,\ell=1}^n =\pm\gamma^\epsilon\det(tr(x_k x_\ell))_{k,\ell=1}^n,\]
where the exponent $\epsilon$ is equal to zero or $n$ according to whether
$i+j$ equals zero or $n$. The former case occurs only once and the latter case
occurs exactly $n-1$ times. The claimed formula then follows.
\end{proof}

\begin{exam}
\label{kaks2ellperhe}
We use the notation from Proposition \ref{syklotomidiskriminantti}. In
\cite{KR} Kiran and Rajan have shown that the family of cyclic algebras
$\A_\ell=(\Q(\zeta_\ell)/\Q(i),\sigma(\zeta_\ell)= \zeta_\ell^5,2+i)$, with
$\ell\ge 3$, consists entirely of division algebras. Let $\Lambda_{nat,\ell}$
be the natural order of the algebra $\A_\ell$. We may now conclude from Lemma
\ref{naturaldiscriminant}, Proposition \ref{syklotomidiskriminantti}, and
Corollary \ref{gausskoppi} that
\[d(\Lambda_{\ell,nat}/\Z[i])=(2+i)^{n(n-1)} (1+i)^{2n^2(\ell-2)},\]
and that
\[m(\Lambda_{nat,\ell})^2=2^{2n^2(\ell-2)} 5^{n(n-1)}.\]
For instance, in the $2$ antenna case $\ell=3, n=2$, we have
$m(\Lambda_{nat,\ell})=80$, and thus the Golden code is denser than the
corresponding lattice $\A_3$ of the same minimum determinant. However, the
natural order of $\A_3$ is not maximal and we will return to this example
later on.
\end{exam}

In Section  \ref{Finding} some facts from the local theory of orders are
required. For the basic properties of localization the reader can turn to
\cite[Chapter 7]{Jac} or \cite[Chapters 1, 2]{R}. For the proofs for the rest
of this section, see \cite{IR} and \cite{Ron}.

If $R$ is a Dedekind domain with a quotient field $F$, and $P$ is a prime
ideal in $R$, then the ring of quotients $R_P=(R/P)^{-1}R\subset F$ is a
discrete valuation ring. For the $R$-lattices $M$ in $\A$  the localization at
$P$ is defined as $M_P=R_PM\subset\A$. $M_P$ is an $R_P$-lattice. Moreover, if
$M$ is a full (cf. Example \ref{left order}) $R$-lattice in $\A$, then $M_P$
is a full $R_P$-lattice in $\A$. To be more specific, let us define the ring
$\Z_p$.

\begin{definition} 
For a rational prime $p$ let $\Z_p$ denote the ring
\[\Z_p=\{\frac{r}{s}\in\Q\ |\ r,s\in\Z,\ gcd(p,s)=1\}.\] 
$\Z_p$ is a discrete valuation ring with the unique maximal ideal $p\Z_p$. If
$\Lambda$ is a $\Z$-order we use the notation $\Lambda_p=\Z_p\Lambda$.
\end{definition}

We remark that one should not confuse the localization $R_P$ with the ring of
integers $\hat R_P$ of the $P$-adic completion. We use the caret to indicate a
complete structure. This is somewhat non-standard in the case of $\Z_p$ that
is nearly universally used to denote the complete ring of $p$-adic integers.
We use $\hat\Z_p$ for the complete ring.

The next statement illustrates a  simple but useful connection between the
orders $\Lambda$ and $\Lambda_p$.

\begin{proposition}
\label{ronyai 2.7} 
Let $\Lambda$ be a $\Z$-order in $\A$. The map
$\Phi:x\mapsto x+p\Lambda_p$, $x\in\Lambda$ induces an isomorphism
of the rings $\Lambda/p\Lambda\cong\Lambda_p/p\Lambda_p$.
\end{proposition}

\begin{proposition}
\label{ronyai 3.1} 
Let $P$ be a prime ideal of the ring $R$. The residue class ring
$\overline{\Lambda}=\Lambda/P\Lambda$ is an algebra with identity element over
the residue class field $\overline{R}=R/P$ and
$dim_F\A=dim_{\overline{R}}\overline{\Lambda}$. If
$\phi:\Lambda\rightarrow\overline{\Lambda}$ is the canonical epimorphism, then
$P\Lambda\subseteq \Rad(\Lambda)=\phi^{-1}\Rad(\overline{\Lambda})$ and $\phi$
induces a ring isomorphism
$\Lambda/\Rad(\Lambda)\cong\overline{\Lambda}/\Rad(\overline{\Lambda})$.
As a consequence, a left (or right) ideal $\I$ of $\Lambda$ is contained in
$\Rad(\Lambda)$ if and only if there exists a positive integer $t$ such that
$\I^t\subseteq P\Lambda$.
\end{proposition}

The following facts establish some practical connections between the local and
global properties of orders.

\begin{proposition}
\label{ronyaialg 2.3} 
Let $\A$ be a simple algebra over $F$. Let $P$ be a prime ideal of $R$, and
$\Gamma$ be an $R$-order in $\A$. Then

(i) $\Gamma_P$ is an $R_P$-order in $\A$.

(ii) $\Gamma$ is a maximal $R$-order in $\A$ if and only if $\Gamma_P$ is a
maximal $R_P$-order in $\A$ for every prime ideal $P$ of $R$.

(iii) $d(\Gamma/R)_P=d(\Gamma_P/R_P)$.
\end{proposition}

\begin{proposition}
\label{ronyaialg 2.5}
Let $P$ be a prime ideal of  $R$ and $\Gamma$ be an $R$-order such that
$\Gamma_P$ is not a maximal $R_P$-order. Then there exists an ideal
$\I\geq P\Gamma$ of $\Gamma$ for which $\OO_l(\I)>\Gamma$.
\end{proposition}

Extremal orders and especially Proposition \ref{ronyai 4.5} below play a key
role in the method for constructing maximal orders.

\begin{definition}
\label{extremal} 
We say that $\Gamma_P$ radically contains $\Lambda_P$ if and only if
$\Lambda_P\subseteq\Gamma_P$ and $\Rad(\Lambda_P)\subseteq \Rad(\Gamma_P)$.
The orders maximal with respect to this partial ordering are called {\it
extremal}. Maximal orders are obviously extremal.
\end{definition}

\begin{proposition}
\label{ronyai 4.1} 
An $R_P$-order $\Lambda_P$ is extremal if and only if
$\Lambda_P=\OO_l(\Rad(\Lambda_P))$.
\end{proposition}

\begin{proposition}
\label{ronyai 4.5} 
Let $\Lambda_P\subset\Gamma_P$ be $R_P$-orders in $\A$. Suppose that
$\Lambda_P$ is extremal and that $\Gamma_P$ is minimal among the $R_P$-orders
properly containing $\Lambda_P$. Then there exists an ideal $\J$ of
$\Lambda_P$ minimal among those containing $\Rad(\Lambda_P)$ such that
$\OO_l(\J)\supseteq\Gamma_P$.
\end{proposition}

\section{Discriminant bound}
\label{DB}

Again let $F$ be an algebraic number field that is finite dimensional over
$\Q$, $\OO_F$ its  ring of integers, $P$ a prime ideal of $\OO_F$ and
$\hat F_P$ the completion. In what follows we discuss the size of ideals of
$\OO_F$. By this we mean that ideals are ordered by the absolute values of
their norms to $\Q$, so e.g. in the case $\OO_F=\Z[i]$ we say that the prime
ideal generated by $2+i$ is smaller than the prime ideal generated by $3$ as
they have norms $5$ and $9$, respectively.

The following relatively deep result from class field theory is the key for
deriving the discriminant bound. Assume that the field $F$ is totally complex.
Then we have the {\em fundamental exact sequence of Brauer groups} (see e.g.
\cite{R} or \cite{M})

\begin{equation}
\label{brauer1}
0\longrightarrow \Br(F)\longrightarrow \oplus \Br(\hat F_P)\longrightarrow
\Q/{\bf Z}\longrightarrow 0.
\end{equation}

Here the first nontrivial map is gotten by mapping the similarity class of a
central division $F$-algebra $\D$ to a vector consisting of the similarity
classes of all the simple algebras $\D_P$ gotten from $\D$ by extending the
scalars from $F$ to $\hat F_P$, where $P$ ranges over all the prime ideals of
$\OO_F$. Observe that $\D_P$ is not necessarily a division algebra, but by
Wedderburn's theorem \cite[p. 203]{Jac} it can be written in the form
\[\D_P=\mathcal{M}_{\kappa_P}(\A_P),\]
where $\A_P$ is a division algebra with a center $\hat F_P$, and $\kappa_P$ is
a natural number called the {\it local capacity}. The second nontrivial map of
the fundamental exact sequence is then simply the sum of the Hasse invariants
of the division algebras $\A_P$ representing elements of the Brauer groups
$\Br(\hat F_P)$.

This  exact sequence tacitly contains the piece of information that for all
but finitely many primes $P$ the resulting algebra $\D_P$ is actually in the
trivial similarity class of $\hat F_P$-algebras. In other words $\D_P$ is
isomorphic to a matrix algebra over $\hat F_P$. More importantly, the sequence
tells us that the sum of the nontrivial Hasse invariants of any central
division algebras must be an integer. Furthermore, this is the only constraint
for the Hasse invariants, i.e. any combination of Hasse invariants $(a/m_P)$
such that only finitely many of them are non-zero, and that they sum up to an
integer, is realized as a collection of the Hasse invariants of some central
division algebra $\D$ over $F$.

Let us now suppose that with a given number field $F$ we would like to produce
a division algebra $\A$ of a given index $n$, having $F$ as its center and the
smallest possible discriminant. We proceed to show that while we cannot give
an explicit description of the algebra $\A$ in all the cases, we can derive an
explicit formula for its discriminant.

\begin{thm}
\label{kokoelma}
Assume that the field $F$ is totally complex and that $P_1,\ldots, P_n$ are
some prime ideals of $\OO_F$. Assume further that a sequence of rational
numbers $a_1/m_{P_1},\ldots , a_n/m_{P_n}$ satisfies
\[\sum_{i=1} ^{n} \frac{a_i}{m_{P_i}} \equiv 0 \pmod{1},\]
$1\leq a_i \leq m_{P_i}$, and $(a_i , m_{P_i})=1$.

Then there exists a central division $F$-algebra $\A$ that has local indices
$m_{P_i}$ and the least common multiple (LCM) of the numbers $\{m_{P_i}\}$ as
an index.

If $\Lambda$ is a maximal $\OO_F$-order in $\A$, then the discriminant of
$\Lambda$ is
\[d(\Lambda/ \OO_F)=\prod_{i=1}^n P_i^{(m_{P_i}-1)\frac{[\A:F]}{m_{P_i}}}.\]
\end{thm}

\begin{proof}
By exactness of the sequence (\ref{brauer1}) we  know that there exists a
central division algebra $\A$ over $F$ which has local indices $m_{P_{i}}$.
From \cite[Theorem 32.19]{R} we know that $\sqrt{[\A:F]}=LCM \{m_{P_i}\}.$
By \cite[Theorem 32.1]{R} the discriminant then equals
\begin{equation}
\label{disc1}
d(\Lambda/R)=\left(\prod_{i=1} ^n {P_i}^{(m_{P_i}
-1)\kappa_{P_i}}\right)^{\sqrt{[\A:F]}},
\end{equation}
where $\kappa_{P_i}$ is the local capacity.

A simple calculation of dimensions shows that
\[\kappa_{P}=\frac{\sqrt{[\A:F]}}{m_{P}}.\]
Substituting this into \eqref{disc1} we get the claim.
\end{proof}

At this point it is clear that the discriminant $d(\Lambda)$ of a division
algebra only depends on its local indices $m_{P_i}$.

Now we have an optimization problem to solve. Given the center $F$ and an
integer $n$ we should decide how to choose the local indices and the Hasse
invariants so that the LCM of the local indices is $n$, the sum of the Hasse
invariants is an integer, and that the resulting discriminant is as small as
possible. We immediately observe that at least two of the Hasse invariants
must be non-integral.

Observe that the exponent $d(P)$ of the prime ideal $P$ in the discriminant
formula
\[d(P)=(m_P-1)\frac{[\A:F]}{m_P}=n^2\left(1-\frac{1}{m_P}\right).\]
As for the nontrivial Hasse invariants $n\ge m_P\ge2$, we see that
$n^2/2\le d(P)\le n(n-1)$. Therefore the nontrivial exponents are roughly of
the same size. E.g. when $n=6$, $d(P)$ will be either $18$, $24$ or $30$
according to whether $m_P$ is $2$, $3$ or $6$. Not surprisingly, it turns out
that the optimal choice is to have only two non-zero Hasse invariants and to
associate these with the two smallest prime ideals of $\OO_F$.

\begin{thm}[Main Theorem]
\label{raja}
Assume that $F$ is a totally complex number field, and that $P_1$ and $P_2$
are the two smallest prime ideals in $\OO_F$. Then the smallest possible
discriminant of all central division algebras over $F$ of index $n$ is
\[(P_1P_2)^{n(n-1)}.\]
\end{thm}

\begin{proof}
By Theorem \ref{kokoelma} the division algebra with Hasse invariants $1/n$
and $(n-1)/n$ at the primes $P_1$ and $P_2$ has the prescribed discriminant,
so we only need to show that this is the smallest possible value.

By the above discussion it is clear that in order to minimize the discriminant
one cannot have more than three nontrivial Hasse invariants. This is because
for prime ideals $P_1,P_2,P_3,P_4$ (listed from the smallest to the largest)
we always have
\[P_1^{d(P_1)}P_2^{d(P_2)}P_3^{d(P_3)}P_4^{d(P_4)}>(P_1 P_2)^{n(n-1)},\]
as the exponents $d(P_i)\ge n^2/2$ irrespective of the values of the Hasse
invariants. A possibility is that some combination of three Hasse invariants
might yield a smaller discriminant. Let us study this in detail.

If one of the local indices, say $m_{P_1}$, has only a single prime factor,
say $p$, then we can add this Hasse invariant together with one of the other
two, as long as we are careful to choose the one, say $m_{P_2}$, whose
denominator is divisible by a smaller power of $p$. In this addition process
the least common multiple of the denominators remains the same, so the new set
of only two nontrivial Hasse invariants corresponds to a division algebra of
the same index. This is because in the sum of the Hasse invariants
\[a_1/m_{P_1}+a_2/m_{P_2}=a'/m'_{P'} \pmod 1\]
the new local index $m'_{P'}$ is gotten from the old local index $m_{P_2}$ by
multiplying it with a (possibly the zeroth) power of $p$. Let $P'$ be smaller
of the two ideals $P_1$ and $P_2$. As $d(P_1)+d(P_2)>n(n-1)\ge d'(P')$, where
$d'(P')$ is the exponent corresponding to the local index $m_{P'}$, this new
division algebra (with nontrivial Hasse invariants associated with primes $P'$
and $P_3$ only) will have a smaller discriminant.

The remaining case is that all the three local indices have at least two
distinct prime factors. In this case all the three Hasse invariants have
numerators $\ge6$. As then $d(P_1)+d(P_2)+d(P_3)>2n(n-1)$, we see that the
discriminant of the division algebra with these Hasse invariants also exceeds
the stated lower bound.
\end{proof}

We remark that in the most interesting (for MIMO) cases $n=2$ and $n=3$, the
proof of Theorem \ref{raja} is more or less an immediate corollary of Theorem
\ref{kokoelma}. We also remark that the division algebra achieving our bound
is by no means unique. E.g. any pair of Hasse invariants $a/n,(n-a)/n$, where
$0<a<n$, and $(a,n)=1$, leads to a division algebra with the same
discriminant.

The smallest primes of the ring $\mathbf{Z}[i]$ are $1+i$ and $2\pm i$. They
have norms $2$ and $5$ respectively. The smallest primes of the ring
$\Z[\omega]$ are $\sqrt{-3}$ and $2$ with respective norms 3 and 4. Together
with Corollaries \ref{gausskoppi} and \ref{eisensteinkoppi} we have arrived at
the following bounds.

\begin{corollary}[Discriminant bound] 
\label{gaussoptimal}
Let $\Lambda$ be an order of a central division algebra of index $n$ over the
field $\Q(i)$. Then the measure of a fundamental parallelotope of the
corresponding lattice
\[m(\Lambda)\ge 10^{n(n-1)/2}.\]
\end{corollary}
\vspace* {4pt}

\begin{corollary}[Discriminant bound] 
\label{eisensteinoptimal}
Let $\Lambda$ be an order of a central division algebra of index $n$ over the
field $\Q(\omega)$, $\omega=(-1+\sqrt{-3})/2$. Then the measure of a
fundamental parallelotope of the corresponding lattice
\[m(\Lambda)\ge (\sqrt{3}/2)^{n^2}12^{n(n-1)/2}.\]
\end{corollary}
\vspace*{4pt}

The Golden algebra reviewed in Example \ref{goldenalgebra} has its nontrivial
Hasse invariants corresponding to the primes $2+i$ and $2-i$ and hence cannot
be an algebra achieving the bound of Theorem \ref{raja}. A clue for finding
the optimal division algebra is hidden in the alternative description of the
Golden algebra given in Example \ref{goldenalgebra}. It turns out that in the
case $F=\Q(i)$, $E=\Q(\zeta)$ instead of using $\gamma'=5$ as in the case of
the Golden algebra we can use its prime factor $\gamma=2+i$.

\begin{proposition}
\label{goldenplusvanha}
The maximal orders of the cyclic division algebra
$\A_3=(\Q(\zeta)/\Q(i),\sigma,2+i)$ of Example \ref{kaks2ellperhe} achieve the
bound of Theorem\ref{raja}.
\end{proposition}

\begin{proof}
The algebra $\A_3$ is generated as a $\Q(i)$-algebra by the elements $\zeta$
and $u$ subject to the relations $\zeta^2=i$, $u^2=2+i$, and $u\zeta=-\zeta
u$. The natural order $\Z[\zeta]\oplus u \Z[\zeta]$ is not maximal. Let us use
the matrix representation of $\A_3$ as $2\times2$ matrices with entries in
$\Q(\zeta)$, so elements of $\Q(i)$ are mapped to scalar matrices and $\zeta$
is mapped to a diagonal matrix with diagonal elements $\zeta$ and $-\zeta$. We
observe that the matrix
\[w=\frac{1}{4}\begin{pmatrix}
2i-(1-i)\sqrt{2}& (2+i)(2i-(1+i)\sqrt{2})\\
(1+i)(1+\sqrt{2}+i)& 2i+(1-i)\sqrt{2}
\end{pmatrix}\]
is an element of $\A_3$. Straightforward calculations show that $w$ satisfies
the equations
\[w^2=-i+iw\quad\hbox{\rm and}\quad w\zeta=-1+\zeta^3-\zeta w.\]
From these relations it is obvious that the free $\Z[\zeta]$-module with basis
elements $1$ and $w$ is an order $\Lambda$. Another straightforward
computation shows that $d(\Lambda/\Z[i])=-8+6i=(1+i)^2(2+i)^2$. As this is the
bound of Theorem \ref{raja} we may conclude that $\Lambda$ is a maximal order.
\end{proof}

By Corollary \ref{gausskoppi} we see that the fundamental parallelotope of the
maximal order in Proposition \ref{goldenplusvanha} has measure $10$. Thus this
code has $2.5$ times the density of the Golden code.

The algebra $\A_3$ has the drawback that the parameter $\gamma$ is quite
large. This leads to an antenna power imbalance in both space and time
domains. To some extent these problems can be alleviated by conjugating the
matrix lattice by a suitable diagonal matrix (a trick used in at least
\cite{WX}). One of the motifs underlying the perfect codes \cite{BORV} is the
requirement that the variable $\gamma$ should have a unit modulus. To meet
this requirement we proceed to give a different construction for this algebra.

\begin{thm}
\label{goldenplus}
Let $\lambda$ be the square root of the complex number $2+i$ belonging to the
first quadrant of the complex plane. The cyclic algebra $\mathcal{GA}+=
(\Q(\lambda)/\Q(i),\sigma,i)$, where the automorphism $\sigma$ is determined
by $\sigma(\lambda)=-\lambda$, is a division algebra. The maximal orders of
$\mathcal{GA}+$ achieve the bound of Theorem \ref{raja}. Furthermore, the
algebras $\mathcal{GA}+$ and $\A_3$ of Theorem \ref{goldenplusvanha} are
isomorphic.
\end{thm}

\begin{proof}
The algebra $\mathcal{GA}+$ is a central algebra $F\{u',\lambda\}$ over the
field $F=\Q(i)$ defined by the relations $\lambda^2=2+i$, $u'^2=i$,
$u'\lambda=-\lambda u'$. Comparing these relations with the relations in the
proof of Theorem \ref{goldenplusvanha} we get an isomorphism of $F$-algebras
$f:\mathcal{GA}+\rightarrow \A_3$ by declaring $f(u')=\zeta$, $f(\lambda)=u$
and extending this in the natural way. The other claims follow immediately
from this isomorphism and Theorem \ref{goldenplusvanha}.
\end{proof}

We refer to the algebra $\mathcal{GA}+$ as the {\it Golden+ algebra}. This is
partly motivated by the higher density and partly by the close relation
between the algebra $\A_3$ and the Golden algebra. After all, the algebra
$\A_3$ comes out when in the alternative description of the Golden algebra
(cf. Example \ref{goldenalgebra}) the variable $\gamma=5$ is replaced with its
prime factor $2+i$. In Section \ref{Finding} we will provide an alternative
proof for Theorem \ref{goldenplus} by explicitly producing a maximal order
within $\mathcal{GA}+$ and verifying that it has the prescribed discriminant.
It is immediate from the discussion in the early parts of this section that in
this case there is only one cyclic division algebra (up to isomorphism) with
that discriminant.

It turns out that all the algebras $\A_\ell$ in the Kiran--Rajan family of
Example \ref{kaks2ellperhe} have maximal orders achieving the discriminant
bound. The following observation is the key to prove this.

\begin{lemma}
\label{kaksitekijainendiskriminantti}
Let $F$ be either one of the fields $\Q(i)$ or $\Q(\omega)$, and let $P_1$ and
$P_2$ be the two smallest ideals of its ring of integers $R$. Let $\D$ be a
central division algebra over $F$, and let $\Lambda$ be any $R$-order in $\D$.
If the discriminant $d(\Lambda)$ is divisible by no prime other than $P_1$ and
$P_2$, then any maximal order $\Gamma$ of $\D$ achieves the discriminant bound
of Theorem \ref{raja}.
\end{lemma}

\begin{proof}
We know that there exists a maximal order, say $\Gamma_0$ containing
$\Lambda$. The discriminant of $\Gamma_0$ is then a factor of $d(\Lambda)$, so
$P_1$ and $P_2$ are the only prime divisors of $d(\Gamma_0)$. From Theorem
\ref{kokoelma} we infer that the only nontrivial Hasse invariants of $\D$
occur at $P_1$ and $P_2$. As the sum of the two Hasse invariants is an
integer, they have the same denominator. This must then be equal to the index
of $\D$. The discriminant formula of Theorem \ref{kokoelma} then shows that
$d(\Gamma_0)$ equals the discriminant bound. Any other maximal order in $\D$
shares its discriminant with $\Gamma_0$.
\end{proof}

\begin{corollary}
\label{optimiperhe}
Let $\ell>2$ be an integer. The maximal orders of the cyclic division algebra
$\A_\ell=(\Q(\zeta_\ell)/\Q(i),\sigma,2+i)$ from Example \ref{kaks2ellperhe}
achieve the discriminant bound.
\end{corollary}

\begin{proof}
Proposition \ref{syklotomidiskriminantti} and Lemma \ref{naturaldiscriminant}
indicate that the only prime factors of the discriminant of the natural order
in $\A_\ell$ are $1+i$ and $2+i$. The claim then follows from Lemma
\ref{kaksitekijainendiskriminantti}.
\end{proof}

At this point we remark that the natural orders of the algebras $\A_\ell$ of
Example \ref{kaks2ellperhe} are very far from being maximal. We will study
this in greater detail in Section \ref{Finding}.

\begin{exam}
\label{eisensteinesimerkki}
Let $F=\Q(\sqrt{-3})$, so $\OO_F=\Z[\omega]$. In this case the two smallest
prime ideals are generated by $2$ and $1-\omega$ and they have norms $4$ and
$3$, respectively. By Theorem \ref{raja} the minimal discriminant is
$4(1-\omega)^2$ when $n=2$. As the absolute value of $1-\omega$ is $\sqrt{3}$
an application of the formula in Corollary \ref{eisensteinkoppi} shows that
the lattice $\mathbf{L}$ of the code achieving this bound has
$m(\mathbf{L})=27/4$. In \cite{HLRV} we showed that a maximal order of the
cyclic algebra $(E/F,\sigma(i)=-i,\gamma=\sqrt{-3})$, where
$E=\Q(i,\sqrt{-3})$, achieves this bound.
\end{exam}

We remark that one of the codes suggested in \cite{WX} is the natural order
of the algebra of Example \ref{eisensteinesimerkki}. However, the authors
there never mentioned the possibility of using a maximal order. Nor did they
mention that their lattice actually is an order.

\section{Finding maximal orders}
\label{Finding}

Consider again the family of cyclic division algebras $\A_\ell$ of index
$n=2^{\ell-2}$ from Example \ref{kaks2ellperhe}. If $\Lambda_{\ell}$ is a
maximal order of $\A_\ell$, then  according to Corollary \ref{optimiperhe}
\[d(\Lambda_{\ell}/\Z[i])=(1+i)^{n(n-1)}(2+i)^{n(n-1)}.\]
On the other hand, by Example \ref{kaks2ellperhe} we know that 
\[d(\Lambda_{\ell,nat}/\Z[i])= (1+i)^{2n^2(\ell-2)}(2+i)^{n(n-1)}.\]
Hence, by Lemma \ref{diskriminanttiindeksi} we may conclude that the natural
order is of index
\[ [\Lambda_{\ell}:\Lambda_{\ell,nat}]=2^{((2\ell-5)n+1)n/2}.\]
In the cases $\ell=3,4,5$ this index thus equals $2^3$, $2^{26}$, and
$2^{164}$, respectively. In other words, using a maximal order as opposed to
the natural order one can send $1.5$, $6.5$, or $20.5$ more bits per channel
use without compromising neither the transmission power nor the minimum
determinant in the respective cases of $2$, $4$, or $8$ antennas! Hence the
problem of actually finding these maximal orders rather than simply knowing
that they exist becomes quite relevant. In the following we shortly depict how
maximal orders can be constructed in general. A more detailed version of the
algorithm can be found in \cite{IR}.

Let again $F$ be an algebraic number field, $\A$ a finite dimensional central
simple algebra over $F$, and $\Lambda$ be a $\Z$-order in $\A$. Assume that
$\A$ is given by relations (e.g. $u^2=\gamma $), and that $\Lambda$ is given
by a $\Z$-basis. For instance, we can always start with the natural order
$\Lambda$ (cf. Example \ref{naturalorder}). We form a set $S=\{p_1,...,p_r\}$
consisting of the rational primes dividing $d(\Lambda)$, i.e. $\Lambda_p$ is a
maximal $\Z_p$-order if $p\notin S$.

The basic idea of the algorithm is to test for $i=1,...,r$ whether $\Lambda$
is maximal at $p_i$. If the answer is yes, $\Lambda$ is a maximal $\Z$-order.
If not, then at the first index $i$ for which $\Lambda_{p_i}$ is not maximal
we can construct a $\Z$-order $\Gamma$ in $\A$ such that
$\Lambda_{p_i}\subset\Gamma_{p_i}$, and hence $\Lambda\subset\Gamma$ (cf.
Propostitions \ref{ronyai 2.7}--\ref{ronyai 4.5}). This can basically be done
in two steps. Let $p\in S$.

\textbf{STEP 1}\ \  REPEAT UNTIL ``YES'':
  Compute $\I=\phi^{-1}(\Rad(\Lambda_p))\leq\Lambda$.
  Does the equality $\OO_l(\I)=\Lambda$ hold?

\hspace{5mm} ``NO'': $\OO_l(\I)\supset\Lambda$

\hspace{16mm} $\Lambda\leftarrow\OO_l(\I)$ (Iteration step)

\textbf{STEP 2}\ \  REPEAT UNTIL ``NO'':
  Compute the minimal ideals $\J_1,\J_2,...,\J_h$ ($h<dim_{\Q}\A$) of
  $\Lambda/p\Lambda$ which contain $\Rad(\Lambda/p\Lambda)$.
  FOR $i=1,...,h$ compute $\I_i=\phi^{-1}(\J_i)$. Does there exist an
  index $i$ for which $\OO_l(\I_i)>\Lambda$?

\hspace{5mm} ``YES'': $\Lambda\leftarrow\OO_l(\I_i)$  (Iteration step)

\hspace{5mm} ``NO'': OUTPUT $\Lambda$ is a maximal $\Z$-order.

Let $p\in S$. First we test whether $\Lambda_p$ is an extremal (cf. Definition
\ref{extremal}) $\Z_p$-order by checking if
$\OO_l(\Rad(\Lambda_p))=\Lambda_p$. If not, then we shall construct a
$\Z$-order $\Gamma>\Lambda$. If $\Lambda_p$ passes this test, then we can use
the test of Proposition \ref{ronyai 4.5}. If there exists an ideal $\J$
minimal among the ideals properly containing $\Rad(\Lambda_p)$ such that
$\OO_l(\J)>\Lambda_p$, then we construct a $\Z$-order $\Gamma>\Lambda$.
Otherwise we correctly conclude that $\Lambda$ is maximal at $p$ and continue
with the next $p$ in the list $S$. In the end, the algorithm yields a
$\Z$-order $\Lambda$ which is now maximal. The algorithm can be used similarly
for constructing $\OO_F$-orders, but in the MAGMA software the implementations
are for $\Z$-orders only.

For more details concerning the computation of the prime ideals in a ring, see
\cite{Ron}. A thorough explanation and an algorithm for computing the radical
can be found in \cite{Ron1}.

Let us next exemplify the above algorithm.

\subsection{$2\times 2$ construction over $\Z[i]$}
\label{2x2}

In the Golden division algebra (cf. Example \ref{goldenalgebra} or
\cite{BRV}), i.e. the cyclic algebra $\mathcal{GA}=(E/F,\sigma,\gamma)$
gotten from the data  $E=\Q(i,\sqrt5)$, $F=\Q(i)$, $\gamma=i$, $n=2$,
$\sigma(\sqrt5)=-\sqrt5$, the natural order $\Lambda$ is already maximal. The
norm of the discriminant of $\Lambda$ (with respect to $\Q$) is $625$, whereas
the norm of the minimal discriminant is 100 \cite{HLRV}. We will now present a
code constructed from a maximal order of the cyclic division algebra
$\mathcal{GA+}$ of Theorem \ref{goldenplus}. The maximal order of
$\mathcal{GA+}$ also admits the minimal discriminant and is in that sense
optimal. The algorithm now proceeds as follows.

The natural order of the algebra $\mathcal{GA+}$ is
$\Lambda=\Z[i]\oplus u'\Z[i]\oplus \lambda\Z[i]\oplus u' \lambda\Z[i]$.
Hereafter, we will  use a shorter notation $\Lambda=\langle 1,u',\lambda,u'
\lambda\rangle_{\Z[i]}$ for this. Let us consider $\Lambda$ at the place
$P=1+i$ as it is the only factor of the discriminant for which we can enlarge
$\Lambda$. The inverse image of the radical (\ref{ronyai 3.1}) is
$\J=\phi^{-1}(\Rad(\Lambda/P\Lambda))=\phi^{-1}(\langle
1+u',1+\lambda,1+u' \lambda\rangle_{\Z_2})=\langle
1+i,1+u',1+\lambda,1+u' \lambda\rangle_{\Z[i]}\subset \Lambda $. 
A straightforward computation shows us (cf. Proposition
\ref{ronyaialg 3.2}) that the element 
\[\rho=\frac{1+u'+\lambda+u' \lambda}{1+i}=\frac{(1+u')(1+\lambda)}{1+i}\in
\OO_l(\J),\]
which means that the answer to the question in Step 1 is ``NO'', and hence we
set $\Lambda'=\langle 1,u',\lambda,\rho\rangle_{\Z[i]}$ and iterate. This
time the inverse image of the radical is
$\J'=\phi^{-1}(\Rad(\Lambda'/P\Lambda'))=\phi^{-1}(\langle
1+u',1+\lambda,1+\rho\rangle_{\Z_2})=\langle 1+i,1+u',1+\lambda,1+\rho
\rangle_{ \Z[i]}\subset \Lambda' $. By taking the element
\[\tau=\frac{u'+\lambda}{1+i}\in\OO_l(\J')\] 
we can again enlarge the order $\Lambda'$ to $\Lambda''=\langle
1,u',\tau,\rho\rangle_{\Z[i]}$ and compute
$\J''=\phi^{-1}(\Rad(\Lambda''/P\Lambda''))=\phi^{-1}(\langle
1+u',\tau,1+\rho\rangle_{\Z_2})=\langle
1+i,1+u',\tau,1+\rho\rangle_{\Z[i]}\subset \Lambda'' $. We need
one more iteration of Step 1. Now the element
\[\nu=\frac{(1+u')(u'+\lambda)}{2}\in\OO_l(\J'')\]
and the order $\Lambda''$ is enlarged to $\Lambda'''=\langle
1,\nu,\tau,\rho\rangle_{\Z[i]}$.  From this iteration we finally get the
answer to be ``YES''.

In Step 2 there is nothing to do, as the only minimal ideal properly
containing the radical is the radical itself. Hence we have constructed a
maximal $\Z[i]$-order of $\mathcal{GA+}$ with a $\Z[i]$-basis
$\{1,\nu,\tau,\rho\}$.

In order to give a concrete description of this order we describe it in terms
of its $\Z[i]$-basis. Let us again denote by $\lambda$ the first quadrant
square root of $2+i$. The maximal order $\Lambda$ consists of the matrices
$aM_1+bM_2+cM_3+dM_4$, where $a,b,c,d$ are arbitrary Gaussian integers and
$M_i, i=1,2,3,4$ are the following matrices.
\begin{align*}
M_1&=\begin{pmatrix}
1 & 0\\
0 & 1\\
\end{pmatrix},\quad
M_2=\frac{1}{2}\begin{pmatrix}
i+\lambda & i-i\lambda\\
1+\lambda & i-\lambda\\
\end{pmatrix},\quad
M_3=\frac{1}{2}
\begin{pmatrix}
(1-i)\lambda & 1+i\\
1-i & (i-1)\lambda\\
\end{pmatrix},\\
M_4&=\frac{1}{2}\begin{pmatrix}
(1-i)(1+\lambda) & (1+i)(1-\lambda)\\
(1-i)(1+\lambda) & (1-i)(1-\lambda) \\
\end{pmatrix}.\\
\end{align*}

\subsection{Enhancements to the Ivanyos--R\'onyai algorithm in some special
cases}

The memory requirements of the above algorithm grow quite rapidly as a
function of the dimension of the algebra. E.g. the MAGMA-implementation runs
out of memory on a typical modern PC, when given the index $8$ cyclic algebra
$\A_5$ of Example \ref{kaks2ellperhe} as an input.

In this subsection we describe an algorithm that finds maximal orders for the
algebras $\A_\ell$. It is an adaptation of the Ivanyos--R\'onyai algorithm
that utilizes several facts special to this family of algebras. We list these
simple facts in the following lemmas. We will denote $\Z[\zeta_\ell]$ by $\OO$
for short.

\begin{lemma}
\label{paaihanne}
The only prime ideal of $\OO$ that lies above the prime 2 is the principal
ideal $P_\ell$ generated by $1-\zeta_\ell$.
\end{lemma}

\begin{lemma}
\label{vapaus}
Let $M$ be a finitely generated free ${\cal O}$-module of rank $k$, and let
and  $m_1,\ldots,m_k$ be a basis. Let $N$ be a submodule of $M$ such that the
index $[M:N]$ is a power of two (in particular this index is finite). Then $N$
is also a free $\OO$-module of rank $k$, and we can find a basis of $N$
of the form
\[n_i=\sum_{j\le i} a_{ij} m_j\quad, a_{ij}\in \OO.\]
\end{lemma}

\begin{proof}
This is a straightforward modification of the proof of the corresponding
result for modules over a PID. We briefly outline the argument, as we will
need this later on. Let us start by choosing a basis $m_1,\ldots,m_k$ for $M$.
We first consider the ideal $I$ of those coefficients of $m_k$ that appear in
expansions of elements of $N$. We have a natural surjective homomorphism from
$M/N$ onto $\OO/I$. Therefore the index of $I$ in $\OO$ is a power of two, so
we may conclude that $I$ is a power of the prime ideal $P_\ell$. By Lemma
\ref{paaihanne} $I$ is a principal ideal generated by a single element
$y_k\in\OO$. We may thus choose an element $n_k=y_k m_k+\sum_{i<k}x_im_i$ from
the submodule $N$. This will be the last element of a basis of $N$. We proceed
by considering the submodule $N'=N\cap \sum_{i<k}\OO m_i$ of vectors whose
last coordinate vanishes. Then any element $n\in N$ can be written in the form
$n=z_kn_k+n'$ where $n'\in N'$. The coefficients of $m_{k-1}$ that appear in
$N'$ then again form an ideal that by the Jordan--Hölder theorem must be a
power of $P_{\ell}$, and the argument can be repeated. In the end we get a
free $\OO$-basis $n_1,n_2,\ldots,n_k$ of $N$ such that
\[n_i=\sum_j b_{ij}m_j,\]
where all the coefficients $b_{ij}\in\OO$, and $b_{ij}=0$ whenever $j>i$.
\end{proof}

\begin{corollary}
The maximal order $\Lambda_\ell$ of $\A_\ell$ is a free $\OO$-module of rank
$n=2^{\ell-2}$.
\end{corollary}

\begin{proof}
We already know that $\Lambda_\ell$ contains $\Lambda_{\ell,nat}$ as a
submodule of a finite index. Thus, there exists an integer $M>0$ with the
property that $M\Lambda_\ell$ is a submodule of finite index in
$\Lambda_{\ell,nat}$. The formula for the discriminants tells us that we can
further select the multiplier $M$ to be a power of two. Clearly, it suffices
to prove that $M\Lambda_\ell$ is a free module of the right rank. As the
natural order, obviously, is a free $\OO$-module of rank $n$, this is a
consequence of Lemma \ref{vapaus}.
\end{proof}

Let then $\Gamma$ be any {\it intermediate order}, i.e. any order with the
property $\Lambda_{\ell,nat}\subseteq\Gamma\subseteq \Lambda_\ell$. We will
denote by $\Gamma_2$ the ring gotten by localizing $\Gamma$ at the prime
$1+i$. This is naturally a subring of the corresponding localized version of
the maximal order and consequently also of the completion of the maximal order
$\hat\Lambda_\ell$. This latter ring is a $\Z_2[i]$-order in the completion of
the central simple $\Q_2(i)$-algebra $\hat\A_\ell$ gotten from $\A_\ell$ by
extending its scalars to the complete field $\Q_2(i)$. Because the algebra
$\A_\ell$ has a full local index $2^{\ell-2}$ at the prime $1+i$,
$\hat\A_\ell$ is actually a division algebra. By \cite[Theorem 12.8]{R}
and the surrounding discussion therein we know that $\hat\Lambda_\ell$ is a
non-commutative discrete valuation ring, and that the $(1+i)$-adic valuation
of the reduced norm serves as a valuation. E.g. it yields a metric subject to
the non-archimedean triangle inequality. So in the matrix representation the
valuation of the determinant distinguishes the units from the non-units in the
ring $\hat\Lambda_\ell$. We immediately see that the same then holds in the
ring $\Gamma_2$ --- the units are precisely the elements whose reduced norm is
a $(1+i)$-adic unit. By the non-archimedean triangle inequality the non-units
of $\Gamma_2$ then form its unique maximal ideal, which is then also the
radical $\Rad (\Gamma_2)$.

We summarize this line of reasoning in the following Lemma that
is the key to our modifications to Step 1 in the main algorithm.

\begin{lemma}
\label{radikaali}
Let $\Gamma$ be any intermediate order. The ideal $\I=\Gamma\cap
\Rad(\Gamma_2)$ consists of exactly those matrices which determinants
are divisible by $1+i$.
\end{lemma}

The following lemma is a simple reformulation of the fact that $P_\ell$ is of
index 2 in $\OO$. It will allow us to reduce the range of certain searches
from $\OO$ to the set $\{0,1\}$.

\begin{lemma}
\label{reduktio}
Assume that $p(x)=\sum_{i=0}^k p_i x^i\in\Z[x]$. Then
\[p(\zeta_\ell)\equiv p_0+p_1+\cdots+p_k \pmod{P_\ell}.\]
\end{lemma}

\vspace*{4pt}
Let us denote by $s_\ell$ the complex number
\[s_\ell=\frac{1}{1-\zeta_\ell}=\frac{1+i}{2}
\left(1+\zeta_\ell+\zeta_\ell^2+\cdots+\zeta_\ell^{n-1}\right).\]
The fractional ideal generated by $s_\ell$ is then $P_\ell^{-1}$.

\begin{proposition}
Let $\Gamma$ be an intermediate order. Assume that it is a free $\OO$-module,
and that $g_1,g_2,\ldots,g_n$ is its basis. Let $I=\phi^{-1}(\Rad(\Gamma_2))$
(cf. Step 1). Then $I$ is also a free $\OO$-module of rank $n$ that
satisfies $\Gamma\subseteq s_\ell I$. We can find a basis for $I$ that is of
the form $r_1,r_2,\ldots,r_n$, where for all $i$ either
\[r_i=g_i+\sum_{j<i}\epsilon_{ij}g_j,\]
such that all the coefficients $\epsilon_{ij}\in\{0,1\}$, or
\[r_i=(1-\zeta_\ell)g_i.\]
\end{proposition}
\vspace*{4pt}

\begin{proof}
Any element of $\Gamma$ has determinant (= its reduced norm) in $\Z[i]$. The
reduced norm of $1-\zeta_\ell$ is an associate of $1+i$. Therefore
$(1-\zeta_\ell)\Gamma\subseteq I\subseteq\Gamma$. Thus the index of $I$ in
$\Gamma$ is a power of two. Hence Lemma \ref{vapaus} implies that $I$ is a
free $\OO$-module of rank $n$. With the notation of Lemma \ref{vapaus} we also
see that the coefficient $y_n$ is always either $1$ or $1-\zeta_\ell$. In the
former case Lemma \ref{reduktio} and the fact that $2\in P_\ell$ allow us to
choose the coefficients $\epsilon_{ij}$ as required. In the latter case we
have no reason not to choose $r_i=(1-\zeta_\ell)g_i$ as this element is in $I$
by Lemma \ref{radikaali}.
\end{proof}

\begin{proposition}
Let $\Gamma$, $I$, and the bases $g_1,\ldots,g_n$ and $r_1,\ldots,r_n$ be as
in the previous proposition. Then the left order $\Gamma'=\OO_\ell(I)$ is a
free $\OO$-module contained in $s_\ell\Gamma$. It has a basis
$g'_1,\ldots,g'_n$, where for all $i$ either
\[g'_i=s_\ell(g_i+\sum_{j<i}\epsilon_{ij}g_j),\]
such that all the coefficients $\epsilon_{ij}\in\{0,1\}$, or
\[g'_i= g_i.\]
\end{proposition}
\vspace*{4pt}

\begin{proof}
The inclusion $(1-\zeta_\ell)\Gamma\subseteq I$ immediately shows that
$\Gamma\subseteq\OO_\ell(I)\subseteq s_\ell\Gamma$. Therefore the index of
$(1-\zeta_\ell)\Gamma'$ in $\Gamma$ is a power of two. Again Lemma
\ref{vapaus} shows that $\Gamma'$ is a free $\OO$-module. We also have the
inclusion $(1-\zeta_\ell)\Gamma'\subseteq\Gamma$. An argument similar to the
one in the proof of the previous proposition then shows that the algorithm in
the proof of Lemma \ref{vapaus} yields a basis of the prescribed type.
\end{proof}

When we use the natural order of the algebra $\A_\ell$ as a starting point, it
is clear that $p=1+i$ is the only interesting prime in Step 1 of the main
algorithm. This step can now be completed simply by letting $\Gamma$ to be the
natural order, and $g_1,\ldots,g_n$ to be its $\OO$-module basis. We next find
a basis for $\Rad(\Gamma)$ by testing, whether any element of the type
$r_i=g_i+\sum_{j<i}\epsilon_{ij}g_j$ has a determinant divisible by $1+i$
(and if no such element is found then including $r_i=(1-\zeta_\ell)g_i$ into
the basis instead). We then proceed to compute an $\OO$-module basis for the
left order $\Gamma'$ of this $\Rad(\Gamma)$. Again we simply check, whether
any elements of the form $g'_i=s_\ell(g_i+\sum_{j<i}\epsilon_{ij}g_j)$ belong
to $\Gamma'$. Observe that it suffices to test a candidate of this form
against the basis elements $r_i$ only. If such an element is found, we record
that $\Gamma'$ will be strictly larger than $\Gamma$. If no such element is
found, we use $g'_i=g_i$ instead. After we have done this for all $i$, we will
know, whether $\Gamma'=\Gamma$. If this is the case, we are done. Otherwise we
replace $\Gamma$ with $\Gamma'$ and repeat the process.

We implemented this on the computer algebra system Mathematica, and on a
typical modern PC it found a maximal order in the case $\ell=5$ in less than
half an hour. We believe that the memory savings due to the use of
$\OO$-bases as opposed to $\Z$-bases in the general purpose implementation
in MAGMA account for this enhancement in the performance of the algorithm.
This algorithm could naturally be ported into any CAS to handle these very
specific cases.

\begin{exam}
Assume that we have the 4 antenna case $\ell=4$. Let us denote $s=s_\ell$ for
short. In this case the above algorithm yields an order with (left)
$\OO$-basis consisting of the elements $u_1,\ldots,u_4$:
\begin{align*}
u_1 =& 1,\\
u_2 =& (s^2+s^3)+s^3u,\\
u_3 =& (s^4+2s^5+2s^6+s^8+s^{10})+(s^5+s^6)u + s^{10}u^2,\\
u_4 =& (s + s^4 + s^5 + s^8 + s^9 + s^{10} + s^{11} + s^{12} + s^{13})
+ (s^9 + s^{11} + s^{13}) u + (s^{12} + s^{13}) u^2 + s^{13}u^3.
\end{align*}
We observe that the highest powers of $s$ appearing in these basis elements
are $0, 3, 10$, and $13$, respectively. This fits well together with our
earlier calculation showing that the index of the natural order in a maximal
one is $2^{26}$, as $s^{-1}$ generates the prime ideal lying above $2$, and
$0+3+10+13=26$.

It is a basic fact from the theory of the cyclotomic rings of integers that
the conjugate of the element $s$ is of the form $\sigma(s)=u_\sigma s$, where
$u_\sigma$ is a unit of the ring $\Z[\zeta]$. Using this observation
and the relation $u s=\sigma(s) u$ we see that instead of the generator
$u_4$ above we could use the product $u_2u_3$. After all, the $\OO$-module
spanned by these elements is an order, so we can utilize the fact that it is
closed under multiplication.
\end{exam}

\begin{exam}
In the $8$ antenna case $\ell=5$ we get a free $\OO$-module of rank $8$ as a
maximal order. The basis elements $u_1,\ldots,u_8$ are similar linear
combinations of $1,u,u^2,\ldots,u^7$ with coefficients of the form $p(s)$,
where $p(x)\in\Z[x]$ and $s=s_\ell$. In this case the polynomial coefficients
of the various basis elements have maximal degrees $0$, $3$, $10$, $13$, $28$,
$31$, $38$, and $41$. As expected, these degrees sum up to $164$. Taking
advantage of the fact that this module is also a ring we can describe the
elements of the basis by
\begin{align*}
u_1 =& 1,\\
u_2 =& (s^2+s^3)+s^3 u,\\
u_3 =& (s+s^2+s^4+2s^5+2s^6+s^8+s^{10})
+ (s^5+s^6)u + s^{10}u^2,\\
u_4 =& u_2u_3,\\
u_5 =& s + 2s^2 + s^3 + 2s^4 + 5s^5 + 8s^6 + 8s^7 + 3s^8 
   + 5s^9 + 6s^{10} + 5s^{11} \\
   &+ 7s^{12} + 6s^{13} + 7s^{14}
   + 4s^{15} + 5s^{16} + 2s^{18} + 2s^{20} + s^{24} + s^{28}\\
   &+ \bigl(s^5 + 2s^6 + 4s^7 + s^8 + s^9 + s^{10} + 2s^{11} + 2s^{12} 
   + 3s^{13} + 3s^{14} + s^{15} + 3s^{16}\bigr) u \\
   &+ \bigl(s^{11} + 2s^{14} + 2s^{15} + s^{16} + s^{18} + s^{20}\bigr)u^2
   + \bigl(s^{15} + s^{16}\bigr)u^3+s^{28}u^4,\\
u_6 =& u_2 u_5,\\
u_7 =& u_3 u_5,\\
u_8 =& u_2 u_3 u_5.
\end{align*}
\end{exam}

\section{Analysis of the perfect algebras}
\label{Perfect}

In this section we illustrate some computational techniques related to Hasse
invariants and discriminants. We use the algebras underlying the perfect
codes as test cases, because this may provide some additional insight into
these algebras. This section places somewhat higher demands on the readers'
background in algebra and algebraic number theory. It may be skipped if
desired, as our code construction will not depend on the material in this
section.

\begin{proposition}
\label{tensoritulo}
Let  $\D_1= (E_1/F, \sigma_1, \gamma_1 )$ and $\D_2=(E_2/F,\sigma_2,\gamma_2)$
be division algebras that have pairwise prime indices $m_1$ and $m_2$. Then
$\D_1 \otimes \D_2$ is a division algebra with an index $m_1 m_2$. Further,
\[\D_1 \otimes \D_2 \simeq (E_1 E_2/F , \sigma_1 \sigma_2, \gamma_1^{m_2}
\gamma_2^{m_1}),\]
where $\sigma_1 \sigma_2$ is an element of $\Gal(E_1 E_2/F) \simeq
\Gal(E_1/F)\times \Gal(E_2 /F)$.

Let  $P_1$ and $P_2$ be some pair of minimal prime ideals of the field  $F$.
If $\D_1$ and $\D_2$ have minimal discriminants that are only divisible by
$P_1$ and $P_2$, then $\D_1 \otimes \D_2$ has a minimal discriminant that is
only divisible by $P_1$ and $P_2$.
\end{proposition}

\begin{proof}
For the proof of the first two claims we refer the reader to \cite[Theorem 20,
p. 99]{AA}. The only nontrivial Hasse invariants of the division algebras
$\D_1$ and $\D_2$ are those associated with primes $P_1$ and $P_2$. The
mappings in the fundamental exact sequence \eqref{brauer1} are homomorphisms
of groups. Together with the fact that extending scalars to a $P$-adic
completion commutes with the formation of a tensor product shows that the
Hasse invariants of $\D_1\otimes \D_2$ are sums of those of $\D_1$ and $\D_2$.
Hence the discriminant of $\D_1\otimes \D_2$ is only divisible by the prime
ideals $P_1$ and $P_2$. By the proof of Theorem \ref{raja} it is then minimal.
\end{proof}

Suppose we have a finite cyclic extension $E/F$ of algebraic number fields.
Let $P$ be a prime of $F$ and $B$  some prime of $E$ that lies over $P$. We
denote the completion $\hat E_B$ by $\hat E_P$ or $E\cdot \hat F_P$. This
notation is valid in Galois extensions, because the fields $\hat E_B$ are
isomorphic for all primes $B$ that lie over $P$.

\subsection{$2\times 2$ perfect code}

The first perfect algebra is the same as the Golden algebra
$\mathcal{GA}=(E/F,\sigma,\gamma)$, where the extension  $E/F={\bf
Q}(i,\sqrt5)/\Q(i)$ has discriminant $(2+i)(2-i)$. The discriminant of the
natural order is therefore $(2+i)^2 (2-i)^2$. Because the discriminant of the
algebra $\mathcal{GA}$ divides $(2+i)^2 (2-i)^2$ it can have at maximum two
prime divisors $(2+i)$ and $(2-i)$. As a consequence the only Hasse
invariants that can be nontrivial are $h_{(2+i)}$ and $h_{(2-i)}$.

The algebra $\mathcal{GA}$ must have at least two nontrivial Hasse invariants
and therefore $h_{(2+i)}$ and $h_{(2-i)}$ are both nontrivial. Combining  the
equations LCM $[m_{(2+i)},m_{(2-i)}]=2$ and $h_{(2+i)}+h_{(2-i)}=1$ we get
that $h_{(2+i)}= h_{(2-i)}=1/2$. Theorem \ref{kokoelma} states that the
discriminant of $\mathcal{GA}$ is $(2+i)^2 (2-i)^2$. Comparing this to the
discriminant of the natural order we see that the natural order is maximal.

\subsection{$3\times 3$ perfect code}

The underlying algebra of the  $3\times 3$ perfect code is $\PP_3=(E/F,
\sigma, \omega)$, where again $\omega=(-1+\sqrt{-3})/2$, $F=\Q(\omega)$, $E=
\Q(\zeta_7 +\zeta_7^{-1},\omega) $ and $\sigma:
\zeta_7 +\zeta_7^{-1}\longmapsto \zeta_7 ^2 +\zeta_7 ^{-2}$. The algebra
$\PP_3$ has a representation as
\[L\oplus  u\cdot L \oplus u^2 \cdot L\]
where $u^3=\omega$.

The discriminant  of the extension  $E/F$ is $(2+\sqrt{-3})^2
(2-\sqrt{-3})^2=P_1 ^2 P_2 ^2$ and the discriminant of the natural order has
therefore only two prime factors. By Lemma \ref{kaksitekijainendiskriminantti}
the only nontrivial Hasse invariants of $\PP_3$  are  $h_{P_1}$ and $h_{P_2}$.
Because LCM [$m_{P_1}$, $m_{P_2}]=3$. We get that $m_{P_1}=m_{P_2}=3$.

To calculate the  Hasse invariant $h_{P_1}$ we pass to the completion
$\PP_{P_1}=F_{P_1}\otimes \PP_3$. From \cite[Theorem 30.8]{R} we get a
cyclic representation $\PP_{P_1}=(\hat E_{P_1}/\hat F_{P_1}, \sigma_{P_1},
\omega)$, where $\hat E_{P_1}/\hat F_{P_1}$ is a totally ramified extension
and $\sigma_{P_1}$ is the natural extension of the automorphism $\sigma$.
Because the local index $m_{P_1}=3$, we know that $\PP_{P_1}$ is a division
algebra.

Next we try to find another cyclic representation for this algebra so that we
can use the definition of Hasse invariant to calculate the value of $h_{P_1}$.

It is readily verified that the field $\hat F_{P_1}(u)= T_{P_1}\subseteq
\PP_{P_1}$ is a cyclic and totally inert extension of $\hat F_{P_1}$. The
Frobenius automorphism of the extension $\hat T_{P_1}/\hat F_{P_1}$ is defined
by the $(\hat T_{P_1}/\hat F_{P_1}, P_1)(u)=u^7$. The Noether--Skolem Theorem
(\cite[Theorem 7.21]{R}) states that there is an element $x \in \PP_{P_1}$
such that
\begin{equation}\label{ehto}
(\hat T_{P_1}/\hat F_{P_1}, P_1)(a)= x^{-1}ax \quad  \forall a \in \hat
T_{P_1}.
\end{equation}

For an element $x$ to fulfill \eqref{ehto} it is enough to satisfy the
equation $(\hat T_{P_1}/\hat F_{P_1}, P_1)(u)=u^7=xux^{-1}$. By considering
the equation $ux=xu^7=x \omega^2 u$ we see that $x=\zeta_7+\zeta_7^{-1}
+\omega^2(\zeta_7^2 +\zeta_7^{-2})+\omega(\zeta_7^4 + \zeta_7^{-4}) \in L $ is
a suitable element.

We now prove that $x^3$ is an element of $F_{P_1}$, and that $v_{P_1}(x^3)=1$.
The first statement follows from $u\sigma(x^3)=x^3 u=x^2u\omega^2x=u x^3.$
The second statement is obtained from the equation
$v_{P_1}(x^3)=v_{P_1}(N_{E/F}(x))=v_{P_1}(7(2+(\sqrt{-3})\omega)=1$.

Proposition \ref{kiransun} now states that $\B_1=(\hat T_{P_1}/\hat F_{P_1},
(\hat T_{P_1}/\hat F_{P_1}, P_1 ), x^3)$ is a division algebra of index $3$.
By \eqref{ehto} we can consider $\B_1$ as a subset of the algebra $\PP_3$. But
$\B_1$ is a $\hat F_{P_1}$-central division algebra and hence a $9$
dimensional vector space over $\hat F_{P_1}$. From this we can conclude that
$(\hat T_{P_1}/\hat F_{P_1}, (\hat T_{P_1}/\hat F_{P_1}, P_1),x^3)=\PP_{P_1}$.

Lemma \ref{Reiner} now implies that $h_{P_1}=1/3$. Because the sum of the
Hasse invariants has to be an integer, the invariant $h_{P_2}$ is $2/3$.

By considering the local indices we see that the discriminant of the maximal
order is $P_1^6 P_2^6$, that is, equal to the discriminant of the natural
order. Thus, the natural order has to be maximal.

\subsection{$4\times 4$ perfect code}

The division algebra under the $4 \times 4$ perfect code is $\PP_4=(E/F,
\sigma, i)$, where $\Q(i)=F$, $\Q(i,\zeta_{15} + \zeta_{15}^{-1})=E$ and
$\sigma: \zeta_{15} + \zeta_{15}^{-1}\longmapsto\zeta_{15}^2 +
\zeta_{15}^{-2}$.

The extension $E/\Q(i)$ has discriminant $d(E/\Q(i))=(2+i)^3 (2-i)^3 (3)^2$,
and the only Hasse invariants that can be nontrivial are $h_{(3)}, h_{(2+i)}$
and $h_{(2-i)}$. We use  similar methods to those in the  case of $\PP_3$ to
get that $h_{(2+i)}=3/4$ and $h_{(2-i)}=1/4$. The sum $h_{(2-i)}+h_{(2+i)}=1$
and therefore $h_{(3)}$ must be trivial. Further, the local indices reveal
that the discriminant of the algebra is $(2+i)^{12}(2-i)^{12}$. The
discriminant of the natural order on the other hand is 
$(2+i)^{12}(2-i)^{12}(3)^8$. Lemma \ref{diskriminanttiindeksi} tells us that
the index of the natural order in the maximal order is $81$.

\subsection{$6\times 6$ perfect code}

In the $6\times 6$ perfect code construction the center is $F=\Q(\omega)$ and
the maximal subfield $E=K(\theta)$, where $\theta=\zeta_{28}+\zeta_{28}^{-1}$.

In \cite{BORV} where the perfect codes were introduced, the authors gave
the mapping $\sigma_1$ by the equation $\sigma_1:\zeta_{28} +\zeta_{28}^{-1}
\longmapsto \zeta_{28}^2 +\zeta_{28}^{-2}$. Unfortunately, this mapping is not
an $F$-automorphism of the field $E$. We replace $\sigma_1$ with the
automorphism $\sigma$ defined by the equation $\sigma :\zeta_{28}
+\zeta_{28}^{-1}\longmapsto \zeta_{28}^5 +\zeta_{28}^{-5}$. The relative
discriminant of the extension $E/F$ is $(2)^6(2+\sqrt{-3})^5
(2-\sqrt{3})^5=(2)^6 (7)^5$. We denote the resulting algebra by $\PP_6$.

Thus the Hasse invariants of $\PP_6$ that can be nontrivial are
$h_{(2+\sqrt{-3})}$, $h_{(2-\sqrt{-3})}$, and $h_{(2)}$.

Now we are going to present $\PP_6$ as a product of two smaller division
algebras. We first calculate the Hasse invariants of these smaller algebras
and then from these derive the Hasse invariants of $\PP_6$.

Let us first consider the algebra $\B_2=(\Q(\sqrt{7}, \omega)/\Q(\omega),
\sigma_2,-\omega)$. The algebra $\B_2$ is a division algebra with Hasse
invariants $h_{(2-\sqrt{-3})}=h_{(2+\sqrt{-3})}=1/2$. The proof is postponed
until the end of Section \ref{Constructing}.

The algebra $\PP_3=(E/F, \sigma_3, \omega)$ was previously shown to be a
division algebra with Hasse invariants $h_{(2-\sqrt{-3})}=2/3$ and
$h_{(2+\sqrt{-3})}=1/3$. We now consider the algebra $\B_3=(E/F, \sigma_3,
\omega^2)$. By \cite[Theorem 30.4]{R} we have $\PP_3\otimes \B_3 \sim (E/F,
\sigma_3, 1)\simeq M_3 (F)$. This shows that $\PP_3\otimes \B_3$ has trivial
Hasse invariants and therefore the Hasse invariants of $\B_3$ are
$h_{(2-\sqrt{-3})}=1/3$ and $h_{(2+\sqrt{-3})}=2/3$.

If we now consider the algebra $\B_3 \otimes \B_2=$
\[(\Q(\sqrt{7},\omega) \cdot \Q(\zeta_7 + \zeta_7 ^{-1},\omega)/\Q(\omega),
  \sigma_2 \sigma_3, (-\omega)^3 \cdot (\omega^2)^2)\]
it is seen that the corresponding Hasse invariants are $h_{(2-\sqrt{-3})}=1/3
+1/2=5/6$ and  $h_{2+\sqrt{-3}}= 1/2+2/3 \equiv 1/6 \pmod{1}$.

By considering the equation $\sigma_3(\zeta_7 + \zeta_7 ^{-1})=\zeta_7 ^2 +
\zeta_7 ^{-2}= \zeta_7 ^5 + \zeta_7 ^{-5}$ we notice that $\sigma_2 \sigma_3
=\sigma_6$. Combining this and the equation $(-\omega)^3 \cdot
\omega^4=-\omega$ we get that $\B_3 \otimes \B_2\simeq \PP_6$.

The algebra $\PP_6$ has only two nontrivial Hasse invariants that are
$h_{(2+\sqrt{-3})}=5/6$ and $h_{(2-\sqrt{-3})}=1/6$. Whence, the discriminant
of the maximal order is $(2-\sqrt{-3})^{30} (2+\sqrt{-3})^{30}=(7)^{30}$.
The discriminant of the natural order on the other hand is $(2)^{36}
(7)^{30}$. In this case Lemma \ref{diskriminanttiindeksi} tells us that the
perfect lattice is of relatively high index $2^{18}$ in its counterpart within
the maximal order of same minimum determinant.

\section{Constructing division algebras with a minimal discriminant}
\label{Constructing}

We have divided this section into two parts. In the first part we are
concentrating on algebras that have a cyclic representation with a unit
non-norm element $\gamma$.

In the second section we relax the restriction on the size of $\gamma$ and we
give a general construction for $\Q(i)$ and $\Q(\sqrt{-3})$-central division
algebras with a minimal discriminant.

One should note that none of the natural orders of the algebras we shall
construct has a minimal discriminant. This, unfortunately, is not just a
coincidence. In the following we prove that there are no natural orders
reaching the bound of Theorem \ref{raja}.

In the next lemma we use some basic results from the theory of discriminants
and differents. For these results and the notion of different we refer the
reader to \cite[Chapter 3.12]{Koch}.

\begin{lemma}
\label{wildramification}
Suppose we have a Galois extension $E/F$ of degree $n$ and that there are $g$
prime ideals $B_i$ of $E$ lying over the prime $P$ of $F$. If the prime $P$
is wildly ramified in the extension $E/F$, then
\[v_P(d(E/F))\geq n.\]
\end{lemma}

\begin{proof}
Suppose that $D_{E/F}$ is the different of the extension $E/F$. Then it is an
easy exercise in Galois theory to show that $v_{B_i}(D_{E/F})=
v_{B_j}(D_{E/F})$ for every $i$ and $j$. Because $P$ was supposed to be wildly
ramified
\begin{equation}\label{different}
s=v_{B_i}(D_{E/F})\geq e,
\end{equation}
where $e$ is the ramification index of $B_i /P$.

The theory of normal extension states that $efg=n$, where $f$ is the inertial
degree of $B_i /P$. Taking into account this and (\ref{different}) we can
conclude that
\[v_P (d(E/F))= v_{P}(N_{E/F}(D_{E/F}))=sgf \geq  egf=n.\]
\end{proof}

\begin{proposition}
\label{epäluonnollinen1}
Suppose we have a division algebra $\D=(E/\Q(i),\sigma, \gamma)$, where
$E/\Q(i)=n$ and $\gamma$ is an algebraic integer. If $\Lambda$ is the natural
order of the division algebra $\D$, then
\[|d(\Lambda/\OO_{\Q(i)})|> |(2+i)^{n(n-1))}(1+i)^{n(n-1)}|.\]
\end{proposition}

\begin{proof}
The natural order $\Lambda$ is a subset of some maximal order $\Lambda_{max}$
and therefore $|d(\Lambda/\OO_{\Q(i)})|\geq |(2+i)^{n(n-1)}(1+i)^{n(n-1)}|$.
Let us then assume that $|d(\Lambda/\OO_{\Q(i)})|=
|(2+i)^{n(n-1)}(1+i)^{n(n-1)}|$.

According to Lemma \ref{naturaldiscriminant} the only primes that could be
ramified in the extension $E/\Q(i)$ are $(1+i)$, $(2+i)$, and $(2-i)$. Lemma
\ref{wildramification} assures that none of these primes could be wildly
ramified.

One of the main results of the global class field theory \cite[p. 124]{M}
states that there exists a ray class field $C_{(1+i)(2+i)(2-i)}$  that
contains all the cyclic extensions of $\Q(i)$ where $(1+i)$, $(2+i)$, or
$(2-i)$ is tamely ramified.

We can now calculate the degree of the extension $C_{(2+i)(1+i)(2-i)}/\Q(i)$.
By \cite[Theorem 1.5]{M} we have $[C_{(2+i)(1+i)(2-i)}:\Q(i)]=2$, which
implies that $E=C_{(2+i)(1+i)(2-i)}$ and $n=2$.

The ray class fields $C_{(2+i)(1+i)}$ and $C_{(2-i)(1+i)}$ that admit tame
ramification at $(2+i)$ and $(1+i)$ or, at $(2+i)$ and $(1-i)$, respectively,
are both trivial extensions of $\Q(i)$. Hence, both $(2+i)$ and $(2-i)$ are
ramified in $E$ and divide the discriminant of the extension $E/\Q(i)$. The
discriminant of the natural order $\Lambda$ now has to be divisible by at
least $(2+i)^2 (2-i)^2$. This gives us a contradiction.
\end{proof}

\begin{proposition}
\label{epäluonnollinen2}
Suppose we have a division algebra $\D=(E/\Q(\sqrt{-3}),\sigma, \gamma)$,
where $E/\Q(\sqrt{-3})=n$ and $\gamma$ is an algebraic integer. If $\Lambda$
is the natural order of the division algebra $\D$, then
\[|d(\Lambda/\OO_{\Q(\sqrt{-3})})|> |(\sqrt{-3})^{n(n-1))}(2)^{n(n-1)}|.\]
\end{proposition}

\begin{proof}
The proof is similar to that of the previous proposition.
\end{proof}

These considerations reveal that reaching the optimal density of a
code-lattice requires considering maximal orders instead of natural ones.

We give one simple lemma for later use, it is a slight generalization to
\cite[Theorem 1]{KR}. We denote the multiplicative ideal group of  the field
$F$ by $(I_F)^*$.

\begin{lemma}
\label{kiransun}
Let $E$ be a Galois extension of a number field $F$ and let $P$ be a prime
ideal of $\OO_F$ that lies under the prime $B$ of the ring $\OO_E$. If the
inertial degree of $P$ in the extension $E/F$ is $f$ and $\gamma$ is such an
element of $F$ that $(v_P(\gamma), f)=1$, then $\gamma^i$ $\notin$
$N_{E/F}(E)$ for any $i = 1, 2 ,\ldots, f-1$.
\end{lemma}

\begin{proof}
The  ideal norm of $B$ is $N_{E/F}(B)=P^f$, where $f$ is the inertial degree
of $P$ in the extension $E/F$. It is clear that the group $N_{E/F}((I_F )^*)$
is generated by the norms of prime ideals and that $\{N_{E/F}(a)\OO_F \,|\,
a\in E^*\}\subseteq N_{E/F}( I_F )$. Therefore $f| v_P(N_{E/F}(a)\OO_F )$ for
all $a \in E$.
\end{proof}

\subsection{Algebras with a unit $\gamma$}

\subsubsection{Center $\Q(i)$}\label{pienigamma}

\begin{table}[thp]
\caption{$\Q(i)$-central division algebras with a unit $\gamma$}
\label{taulukko1}
\begin{center}
\renewcommand{\arraystretch}{1.44}
 \begin{tabular}{|c|c|l|} \hline
 $n$ & $\gamma$ & $f_{n}$  \\ \hline
 $2$ & $i$      & $x^2 + (2+i)$  \\
 $4$ & $i$      & $x^4 + (2+i) $ \\\hline
\end{tabular}
\end{center}
\end{table}

In Table \ref{taulukko1} we give a cyclic representations for algebras of
degree $2$ and $4$.  Proposition \ref{albert} implies that $4$ is the biggest
degree that we can hope to have a cyclic division algebra with a unit
$\gamma$. There does not exist such an algebra of degree $3$. The reason for
this is that in every cyclic extension $E/\Q(i)$ of degree three, all the
units of $\Q(i)$ are third powers and therefore are in the image of the norm
$N_{E/\Q(i)}$.

In the following we use the generic notation $\Q(i)=F$ and $E=F( a_n )$, where
$a_n$ is a zero of the polynomial $f_n$.

\emph{Algebra $\D_2$:}
The algebra $\D_2$ was previously shown to be a division algebra with a
minimal discriminant.

\emph{Algebra $\D_4$:}
When considering $\D_4$ we first have to check whether it really is a division
algebra. We note that $(2+i)$ is a totally ramified prime in $E /F$. This
results in the local extension $E_{(2+i)}/F_{(2+i)}$ being a totally and
tamely ramified cyclic extension of degree $4$. We note that $\#
(\OO_{{F}_{(2+i)}}/ (2+i)\OO_{{F}_{(2+i)}})=\#(\OO_F/ (2+i))=5$.

Proposition \ref{albert} states that $\D_4$ is a division algebra if $i$
satisfies the norm condition, i.e. neither of the elements $\{i, -1 \}$ is a
norm.

Hasse Norm Theorem \cite[Theorem 32.8]{R} states that it is enough to show
that the elements $\{i, -1\}$ are not norms in the extension $\hat
E_{(2+i)}/\hat F_{(2+i)}$. Elementary local theory \cite[Proposition 7.19]{M2}
states that if we have any complete residue system $\{0,1,a,b,c\}$  of the
group $\OO_{\hat E_{(2+i)}}/(2+i)\OO_{\hat E_{(2+i)}}$ and an arbitrary unit
$e\in \hat F_{(2+i)} $ then
\begin{equation}\label{multiplikat}
\hat E_{2+i}^*=\{1,a,b,c\}\times( 1+ (2+i)\OO_{\hat E_{2+i}})  \times
\langle e(2+i) \rangle.
\end{equation}
The prime $(2+i)$ is tamely ramified in $\hat E_{(2+i)}/\hat F_{(2+i)}$ and
therefore the \emph{local conductor} is $(2+i)$ (\cite[p. 12]{M}). The
definition of the conductor now implies that $( 1+(2+i)\OO_{{\hat E}_{2+i}} 
\subseteq N_{\hat E_{(2+i)}/\hat F_{(2+i)}}(\hat E_{(2+i)})$. Because
the prime $(2+i)$ is totally ramified, we have $e_1 (2+i) \subseteq
N_{\hat E_{(2+i)}/\hat F_{(2+i)}}(\hat F_{(2+i)})$ for some unit $e_1 \in
\hat F_{(2+i)}$. The previous results now imply that $( 1+ (2+i)\OO_{{\hat
E}_{(2+i)}}) \times \langle e_1(2+i) \rangle
\subseteq N_{\hat E_{(2+i)}/\hat F_{(2+i)}}(\hat E_{2+i})$.

One of the main theorems of local class field theory states
that $(\hat F_{(2+i)})^*/(N_{\hat E_{(2+i)}/\hat F_{(2+i)}}(\hat
E_{2+i}^*)) = \Gal(\hat E_{(2+i)}/\hat F_{(2+i)})$. By considering
\eqref{multiplikat} we see that the elements  $\{a, b,c\}$ are not norms.
Because the elements $\{0,i, -1, -i, 1 \}$ form a complete residue system of
the group $\OO_{\hat E_{(2+i)}}/(2+i)\OO_{\hat E_{(2+i)}}$ we find that
neither of the elements $\{i, -1\}$ is a norm.

The discriminant of the extension $E/F$ has only two prime divisors $(2+i)$
and $(1+i)$ and therefore also the discriminant of the natural order of $D_4$
has only two prime divisors. This implies that the discriminant of the algebra
is minimal.

\subsubsection{Center $\Q(\sqrt{-3})$}

\begin{table}[thp]
\caption{$\Q(\omega)$-central division algebras with a unit $\gamma$}
\label{taulukko2}
\renewcommand{\arraystretch}{1.44}
\begin{center}
\begin{tabular}{|l|c|l|} \hline
 $n$ & $\gamma$    & $f_{n}$ \\ \hline
 $2$ & $-\omega$   & $x^2 + \sqrt{-3}$ \\
 $3$ & $\omega$    & $x^3 -2$ \\
 $6$ & $-\omega^2$ & $ x^6 -3\sqrt{-3}x^4 + 4x^3 -9x^2 + 12\sqrt{-3}x +
3\sqrt{-3} + 4$ \\\hline
\end{tabular}
\end{center}
\end{table}
In Table \ref{taulukko2} we give cyclic representations for algebras of
degrees $2$, $3$, and $6$. The theorem of Albert shows that $6$ is the
biggest degree we could hope to have a division algebra with a unit $\gamma$.
We cannot have a division algebras of degrees $4$ and $5$ as tensoring these
with a division algebra $\G_3$ (below) would respectively give us division
algebras of degrees $12$ and $15$ with a unit $\gamma$.

We use the same generic notation as in the case of $\Q(i)$-central algebras.

\emph{Algebra $\G_2$:}
We use here the same methods that were used with the algebra $\D_4$. We remark
that $(\sqrt{-3})=P$ is tamely ramified in the extension $E/F$. If we pass
to the completion $E_{P}/F_{P}$ we get that the local conductor is $P$ and
that $\{-\omega, 1,0 \}$ is a complete set of representatives of the group
$\OO_{{F_P}/P} $. As a result it is seen that $-\omega$ is not a norm in the
extension $E_{P}/F_{P}$ and therefore it is not a norm in the extension $E/F$
either. From this it follows that $\G_2$ is a division algebra.

By now it is obvious that the discriminant of the natural order of the
algebra $\G_2$ has only two divisors $(\sqrt{-3})$, and $(2)$ and hence
the maximal order admits a minimal discriminant.

\emph{Algebra $\G_3$:}
The proof of this case is similar to that of $\G_2$ except that the tamely
ramified prime $P$ is $2$ and that the suitable set of representatives is 
$\{1, \omega, \omega^2 \}$.

\emph{Algebra $\G_6$:}
The algebra $\G_6$ we got as a tensor product from the algebras $\G_2$ and
$\G_3$.

\emph{The postponed proof}.
When we were discussing the $6\times 6$ perfect code we postponed the analysis
of the algebra $\B_2=(E/F, \sigma_2, -\omega)$, where $E/F=\Q(\sqrt{7},
j)/\Q(\omega)$. Now we have enough methods to attack this problem. We use
similar strategy as in the case of the algebra $\D_4$.

The prime  $(2+\sqrt{-3})=P_1$  is tamely ramified in the extension $E/F$. By
passing to the $P_1$-adic completion $\hat E_{P_1}/\hat F_{P_1}$ we find that
the local conductor is $P_1$. The image of the norm $N_{\hat E_{P_1}/\hat
F_{P_1}}$ includes $\langle (1+P_1) \rangle \times \langle e(2+\sqrt{-3})
\rangle$, where $e$ is a unit of $\hat F_{P_1}$.

The  set $\{0,1,\omega, -\omega, \omega^2, -\omega^2 \}$ is a complete residue
system of the group $\OO_{F_{P_1}}/ P_1 \OO_{F_{P_1}}$ and whence
\[(F_{P_1})^* = \langle -j \rangle \times(1+P_1 ) \times
\langle e(2+\sqrt{-3}) \rangle.\]
On the other hand  $\#((F_{P_1})^*/ N_{E_{P_1}/F_{P_1}}(E_{P_1}^*))=2$ and
therefore $-j$ cannot be a norm. From this it follows that the local algebra
$(\B_2)_{P_1}$ is a division algebra of index two.

There is no other choice for the Hasse invariant $h_{P_1}$ than $1/2$.

Replacing the prime $P_1$ with $P_2 =(2-\sqrt{-3})$ in previous
considerations we see that $h_{P_2}=1/2$.

The extension $E/F$ has only  three ramified primes $(2-\sqrt{-3}),
(2+\sqrt{-3})$, and $(2)$. Thus, the discriminant of the algebra $\B_2$
can have three prime divisors at maximum. The potential nontrivial Hasse
invariants of $\B_2$ are now $h_{P_1}$, $h_{P_2}$, and $h_{(2)}$. The sum of
$h_{P_1}$ and $h_{P_2}$ is $1$ and therefore $h_{(2)}$ must be trivial.

\subsection{General construction}
\label{quadratic}

In their recent paper \cite{EKPKL} Elia et al. gave an explicit construction
for division algebras of an arbitrary degree with centers $\Q(i)$ and
$\Q(\sqrt{-3})$. In their general constructions they used non-unit, but
relatively small $\gamma$'s. As they were not interested in maximal orders nor
the discriminants of the corresponding division algebras their algebras (with
few exceptions) did not happen to have minimal discriminants.

We are now going to give a general construction for division algebras of
arbitrary degree and with minimal discriminants. Due to Proposition
\ref{tensoritulo} we can concentrate on algebras of prime power index. As a
drawback our constructions will be dependent on the existence of certain prime
numbers. We discuss this existence problem in Section \ref{existence} which is
purely number theoretic.

We first consider two easy prime powers and then move forward to more
complicated ones.

For ease of notation in this subsection we will denote by $\Z_m$ the residue
class ring modulo $m$, i.e. $\Z_m=\Z/m\Z$. Thus e.g. $\Z_m^*$ is logically the
group of units of that ring.

\begin{lemma}
\label{method1}
Suppose that $E$ is a cyclic extension of $F$ and that $a\OO_E = P_1$ and
$P_2$ are a pair of smallest primes in  $F$. Assume that $P_1$ is totally
inert and $P_2$ is the only ramified prime in the extension $E/F$. Then
\[\A=(E/F, \sigma, a ),\]
where $\langle \sigma\rangle =\Gal(E/F)$, is a division algebra that has a
minimal discriminant.
\end{lemma}

\begin{proof}
Lemma \ref{kiransun} combined with Proposition \ref{albert} gives that $A$
is a division algebra. The minimality of the discriminant follows from Lemma
\ref{kaksitekijainendiskriminantti}.
\end{proof}

\begin{exam}
\label{esimperhe}
Lemma \ref{method1} is nothing but a simple generalization of Corollary
\ref{optimiperhe} where we gave a construction for a family of $\Q(i)$-central
division algebras of degree $2^k$ with a minimal discriminant.
\end{exam}

\begin{exam}
\label{otorni}
The field $\Q(\zeta_{3^{k+1}})$ has a unique subfield $Z$ with $[Z:\Q]=3^k$.
The extension $\Q(\sqrt{-3})Z/$ $\Q(\sqrt{-3})$ has degree $3^k$ and the prime
$(2)$ is totally inert in this extension. The extension also has a very
limited ramification, the prime $(\sqrt{-3})$ is the only ramified one.

Primes $(\sqrt{-3})$ and $(2)$ are a pair of minimal primes in the field
$\Q(\sqrt{-3})$. Lemma \ref{method1} states now that the cyclic algebra
$A=(\Q(\sqrt{-3})Z/\Q(\sqrt{-3}), \sigma, 2)$ is a division algebra with a
minimal discriminant.
\end{exam}

In Example \ref{otorni} we found a suitable extension $E/\Q(\sqrt{-3})$ that
only had one ramified prime ($\sqrt{-3}$). However we can prove that for an
arbitrary degree there usually does not exist a cyclic extension that has
ramification  over ($\sqrt{-3}$) or $(2)$ only. This assures that in general
we cannot use such simple methods. Next we will provide a construction method
that takes care of most of the prime power degrees. First we need some
preliminary results.

We now present a global Frobenius automorphism. Suppose we have a finite
Galois extension $E/F$ and that $B$ is such a prime ideal of $\OO_E$ that
$B\cap\OO_F=P$ is unramified in the extension $E/F$. There exists a unique
element $(B, E/F)$ of the group $\Gal(E/F)$ that is associated to the prime
$B$. We call this element the Frobenius automorphism of $B$.

If the extension $E/F$ is abelian, all the primes $B_i$ that lie over $P$ have
the same Frobenius automorphism and we can denote $(B, E/F)$ by $(P,E/F)$.

For the definition and properties of the Frobenius automorphism we refer
the reader to \cite[p. 379]{Nark}.

We consider a tower of fields $F_1\subseteq F_2\subseteq E$ of finite
extensions.

\begin{proposition}
\label{Frobenius}
If $F_1\subseteq F_2 \subseteq E$, $E/F_1 $ and $F_2/F_1$ are normal and $B$
is such a prime ideal of $E$ that $B\cap F_1=P$ is unramified in $E/F_1$, then
\[(B, E/F_1 )|_{F_2} = (B\cap F_2, F_2 / F_1).\]
The  prime $P$ is totally inert in the extension $E/F_1$ if and only if $(B,
E/F_1)$ generates the group $\Gal(E/F_1)$.
\end{proposition}

\begin{proof}
\cite[Theorem 7.10, p. 380]{Nark}.
\end{proof}

The next lemma is a rather direct consequence of the definition of Hasse
invariant.

\begin{lemma}
\label{Reiner}
Let
\[\A=(E/F, \sigma, \gamma)\]
be a division algebra where $\langle \sigma \rangle =G(E/F)$, $\gamma \in
F^*$, $[E:F]=n$ and suppose that $P$ is  a prime ideal of $F$ that is totally
inert in the extension $E/F$. If $k$ is the smallest possible positive integer
so that $\sigma^k$ is the Frobenius automorphism of $P$ then the Hasse
invariant of $P$
\[h_P= \frac{k v_P(\gamma)}{n}.\]
\end{lemma}

\begin{proof}
\cite[p. 281]{R}.
\end{proof}

Let us next consider a tower of fields $F_1\subseteq F_2\subseteq E$ of finite
extensions and the proofs of the next two simple lemmas will be omitted.

\begin{lemma}
\label{towers}
Let $B$ be a prime ideal of $E$, $P_2=\OO_{F_2}\cap B$ and $P_1=\OO_{F_1}\cap
B$.

1.  Let $f(B/P_1)$, $f(B/P_2)$, and $f(P_2/P_1 )$ be the respective inertia
degrees of $B$ over $P_1$, $B$ over $P_2$, and $P_2$ over $P_1$. Then
\[f(B/P_1)=f(B/P_2)f(P_2/P_1).\]

2. Let $e(B/P_1)$, $e(B/P_2)$, and $e(P_2/P_1 )$ be the respective
ramification indices of $B$ over $P_1$, $B$ over $P_2$, and $P_2$ over $P_1$.
Then
\[e(B/P_1)=e(B/P_2)e(P_2/P_1).\]
\end{lemma}

\begin{lemma}
\label{inra}
Let $E/F$ be a Galois extension, $B$ a prime ideal of $E$  and $P=F\cap B$.
Then
\[e(B/P) \mid [E:F]\]
and
\[f(B/P) \mid [E:F].\]
\end{lemma}

\begin{lemma}
\label{alkuluku}
Let $p$ be a prime and $n$ such an integer that $n|(p-1)$. The field
$\Q(\zeta_p) $ has a unique subfield $Z$ with $[Z:\Q]=n$.

There exists a group isomorphism $\phi$ from $\Z_{p}^*/(\Z_p^*)^n$ to
$\Gal(Z/\Q)$ that takes any prime $p_i\neq p$ to the corresponding Frobenius
automorphism $(p_1, Z/\Q)$ in $\Gal(Z/\Q)$.

The prime $p_1 \neq p$ is totally inert in the extension $Z/\Q$ if and only
if $p_1^t$ is not an $n$th power $\!\pmod p$ for $t=1,\dots,n-1$.
\end{lemma}

\begin{proof}
It is well known that there exists a unique isomorphism $\psi$ from $\Z_p^*$
to $\Gal(\Q(\zeta_p) /\Q)$ which takes prime $p_1\neq p$ to $(p_1, \Q(\zeta_p)
/\Q)$. We denote the fixed field of the group $\psi(\Z_p ^*)^n$ by $Z$. It is
now clear that $Z$ is unique and $[Z:\Q]=n$. If we first map the elements of
$\Z_p^*$ with  $\psi$ to $\Gal(\Q(\zeta_p) /\Q)$ and then restrict the
resulting automorphisms to the field $Z$, we obtain an isomorphism $\phi$
from $\hat\Z_{p} ^* / (\hat\Z_p^*)^n$ to $\Gal(Z/\Q)$. Proposition
\ref{Frobenius} states that $\phi$ has the claimed properties.

The last claim follows from the properties of $\phi$ combined with the last
statement of Proposition \ref{Frobenius}.
\end{proof}

\begin{proposition}\label{simple2}
Suppose that $F=\Q(\sqrt{c})$ is a quadratic field, $q \neq 2$ is a given
prime and $n$ a given integer. We suppose that $P_1$ and $P_2$ are the
smallest primes  ideals in $F$ and $p_1$ and $p_2$ are the prime numbers that
lie under $P_1$ and $P_2$.

Let $p$ be such a prime that $q^n | (p-1)$, $(p,c)=1$ and that $p_1$ and $p_2$
are totally inert in the extension $Z/\Q$, where $Z$ is the unique subfield of
$\Q(\zeta_p)$ of degree $q^n$. We also suppose that $p$ is inert in the
extension $F/\Q$.

The extension $FZ/F$ is a cyclic Galois extension of degree $q^n$ where the
prime ideals $P_1$ and $P_2$ are totally inert and $P=p\OO_F$ is the only
ramified  prime ideal in the extension $FZ/F$.
\end{proposition}

\begin{proof} 
Let $B$ be a prime ideal of $FZ$, $P_Z=\OO_{Z}\cap B$, $P_F= \OO_{F}\cap B$
and $b=\Q \cap B$. We denote the corresponding ramification indices by
$e(B/P_Z)$, $e(P_Z/P_F)$ and $e(P_F/b)$. According to Lemma \ref{towers}
\[e(B/b)=e(B/P_Z)e(P_Z/b)=e(B/P_F)e(P_F/b).\]
Lemma \ref{inra} for its part states that $e(B/P_Z),e(P_F/b) \mid 2$ and
$e(P_Z/b),e(B/P_F)\mid q^n$. This together with the previous equation shows
that the prime $P_F \subset \OO_F$ is ramified in the extension $FZ/F$ if and
only if the prime $b$ is ramified in the extension $Z/\Q$.

The prime $p$ is the only ramified prime in $Z/\Q$ and because $p$ is inert in
the extension $F/\Q$ we see that $P$ is the only ramified ideal in the
extension $ZF/F$.

If  we choose $B$ so that $P_F=P_1$ or $P_F=P_2$, then
\[f(B/b)=f(B/P_Z)f(P_Z/b)=f(B/P_F)f(P_F/b)= q^n \cdot c,\]
where $c=1$ or $c=2$. This combined with Lemma \ref{inra} implies that
$f(B/P_F)=q^n$.
\end{proof}

In the following propositions we use the notation from Proposition
\ref{simple2}.

\begin{proposition}
\label{hunting gamma}
There exists  such a group isomorphism between $\Gal(FZ/F)$ and $\Gal(Z/\Q)$
that every Frobenius automorphism of $B \subset \OO_{FZ}$ maps to the
Frobenius automorphism of $B\cap Z=B_Z $.
\end{proposition}

\begin{proof}
It is a well-known fact that there exists a well defined surjective
homomorphism from $\Gal(FZ/\Q)$ to $\Gal(Z/\Q)$ for which $\sigma\longmapsto
\sigma|_{Z}$. The kernel of this map consists of those elements of
$\Gal(FZ/\Q)$ that act trivially on the field $Z$. On the other hand, if we
restrict the domain of the map to those elements that act trivially on $F$
this map is an injection because the only element of $\Gal(FZ/\Q)$ that acts
trivially on both fields $F$ and $Z$ is the identity map. As we know that
$\left|\Gal(FZ/F)\right|=\left |\Gal(Z/\Q)\right|$ the described map must
be an isomorphism. Now the statement about Frobenius maps follows from
Proposition \ref{Frobenius}.
\end{proof}

\begin{proposition}
\label{gamma}
Let
\begin{equation}\label{yht1}
p_2 p_1 = 1
\end{equation}
in the group $\Z_p ^* /(\Z_p ^*)^{q^n}$, $P_1=a_1\OO_F$, and $P_2=a_2\OO_F$.
Then
\[\A= (FZ/F, \sigma, a_1a_2)\]
with  $\langle \sigma \rangle= \Gal(FZ/F)$ is a division algebra that has a
minimal discriminant.
\end{proposition}

\begin{proof}
The prime  $P_1$ is totally inert in the extension $FZ/F$. Thus, Lemma
\ref{kiransun} states that $\A$ is a division algebra.

From the cyclic presentation of the algebra $\A$ we instantly see that $\A$
has only three Hasse invariants that can be nontrivial: $h_{P_1}$, $h_{P_2}$,
and $h_P$. In what follows we are going to show that the invariant $h_P$
must be trivial.

We first choose  $\sigma$ to be the Frobenius automorphism of $P_1$. Lemma
\ref{Reiner} now shows that the Hasse invariant of $P_1$ is
\[\frac{1}{q^n}=h_{P_1}.\]

Because the group $\Z_p^*/(\Z_p^*)^{q^n}$ is cyclic we get from \eqref{yht1}
that $p_2=p_1^{q^n -1}$ in $\Z_p ^* /(\Z_p ^*)^{q^n}$. This implies that
$(P_2, FZ/ F)=\sigma^{n-1}$. Lemma \ref{Reiner} then states that
\[\frac{q^n -1}{q^n}=h_{P_2}.\]

The sum of the Hasse invariants of $\A$ must be zero $\pmod{1}$, whence
\[h_{P_1} + h_{P_2} + h_{P} \in \Z.\]
But, we already saw that $h_{P_1} + h_{P_2} \in \Z$, which implies that $h_{P}
\in \Z$. The discriminant of the algebra $\A$ has now only two divisors $P_1$
and $P_2$.

In the beginning of our proof we make the assumption that $\sigma$ is the
Frobenius of the prime $P_1$. However, the choice of the generator of the
group $\Gal(FZ/F)$ in a cyclic representation does not change the discriminant
of the corresponding algebra.
\end{proof}

\begin{exam}
\label{etsintaalgoritmi}
Suppose that the center $F=\Q(i)$. The primes $(1+i)$ and $(2+i)$ are a pair
of smallest prime ideals in this field. We want to produce a division algebra
of index $10$ that has a minimal discriminant. It is not difficult to check
that $2^t$ and $5^t$ are not $5$th powers $\!\pmod{11}$ for $t=1,\dots,4$,
and that $11$ is inert in the extension $F/\Q$. Lemma \ref{simple2} states
that $\Q(\zeta_{11})$ has a subfield $Z$, $[Z:\Q]=5$, and that $2$ and $5$ are
totally inert in the extension $Z/ \Q$.

Proposition \ref{simple2} states that the primes $(1+i)$ and $(2+i)$ are
totally inert in the extension $FZ/F$ and the prime ideal $11\OO_F$ is the
only ramified ideal in the extension $FZ/F$.

We easily see that $2 \cdot 5=1$ in $\Z_{11}^*/({\Z_{11}^*})^5$. Therefore,
\[(FZ/F, \sigma_1, (1+i)(2+i))\]
is a division algebra with a minimal discriminant.

We previously saw that $\A=(\Q(\zeta_{2^4})/F, \sigma_2, 2+i)$ is a division
algebra of index $2$ and has a minimal discriminant. Finally, from Proposition
\ref{tensoritulo}
\[(\Q(\zeta_{2^4})Z/ F, \sigma_1 \sigma_2, (1+i)^2 (2+i)^{7})\]
is seen to be a division algebra of degree $10$ with a minimal discriminant.
\end{exam}

\subsection{Existence of suitable primes}
\label{existence}

Propositions \ref{simple2} and \ref{gamma} have turned our construction
project into a hunt of suitable prime numbers. The problem is that we do not
know if there are ``enough'' suitable prime numbers. The answer is that in
most cases there are. This will be proved in Theorem \ref{main}, but first we
need some preliminary results.

For the definition and the basic properties of Kummer extensions we refer the
reader to \cite[p. 197]{Koch}.

\begin{proposition}
\label{kummer}
Let $E/F$ be a Kummer extension with $E=F(\alpha)$, $\alpha^n=a\in \OO_F$, and
let $P$ be a prime ideal of $F$ that is not a divisor of $a \cdot n$.
Furthermore, let $t$ be the largest divisor of $n$ such that the congruence
\[x^t \equiv a \pmod{p}\]
has a solution in $\OO_F$. Then $P$ decomposes in $E$ into a product of $t$
prime ideals of degree $n/t$ over $P$.
\end{proposition}

\begin{proof}
\cite[Theorem 6.8.4, p. 197]{Koch}.
\end{proof}

\begin{lemma}
\label{index}
Suppose that $q$ and $p$ are prime numbers and that $ q^t |(p-1)$ for some
integer $t$. If $c$ is an integer and the equation
\begin{equation}\label{turhuus1}
c \equiv x^q  \pmod{p}
\end{equation}
is not solvable, then neither is any of the equations
\begin{equation}\label{turhuus2}
c^k \equiv x^{q^t} \pmod{p},
\end{equation}
where $k=1,\dots,q^t -1$.
\end{lemma}

\begin{proof}
Let $a$ be a generator of the cyclic group $\Z_p^*$. Then we can write that
$c\equiv a^n \pmod p$ for some integer $n$.

Let us assume that \eqref{turhuus1} has no solution. This implies that $q$ is
not a factor of $n$. Assume then that for some $k$ there is a solution $d$ for
\eqref{turhuus2}. If we write $d \equiv a^s$, then \eqref{turhuus2} gives that
$kn-sq^t=v(p-1)$, where $v$ is some integer. As $q^t | (p-1)$ this would mean
that $q^t | kn$. That gives us a contradiction.
\end{proof}

In the following we use the phrase ``the prime $P$ has inertia in the
extension $E/F$''. By that we mean that at least one prime ideal $B$ of $E$
that lies over the $P$ has inertial degree $f(P|B)>1$.

\begin{lemma}
\label{kompo}
Suppose that $F_1$ and $F_2$ are Galois extensions of a field $F$ and $F_1\cap
F_2 =F$. The prime $P$ of $\OO_F$ has inertia in the extension $F_1 F_2$ if
and only if it has inertia in the extension $F_1$ or $F_2$. The prime $P$ is
ramified in the extension $F_1 F_2$ if and only if it is ramified in $F_1$ or
in $F_2$.
\end{lemma}

\begin{proof}
For the proof the reader is referred to \cite[p. 263]{Rib}.
\end{proof}

The proof of the following theorem is a slightly modified version of the proof
of \cite[Theorem 1]{Perl}. We do not suppose here that the center is totally
complex nor that the ring $\OO_F$ is a PID. However, we suppose that $p_1\neq
p_2$.

\begin{thm}
\label{main}
Assume that $F=\Q(\sqrt{c})$ is a quadratic field, $P_1$ and $P_2$ are the
smallest primes in $F$, $q \neq 2$ is a given prime, and $n$ a given integer.
Let us also suppose that $p_1$ and $p_2$ are prime numbers that lie under
$P_1$ and $P_2$.

If $q\nmid c$, then there exists infinitely many prime numbers $p$ so that $p$
is inert in $F$, $\Q(\zeta_p )$ has a unique subfield $Z$, $[Z:\Q]=q^n$, where
$p_1$ and $p_2$ are totally inert, and $p_1 p_2 =1$ in $\Z_p^*/
(\Z_p^*)^{q^{n}}$.
\end{thm}

\begin{proof}
Let us denote $q^n=s$, $\Q(\zeta_s )(( p_1 p_2 )^{1/s})=K$, $K((p_1)^{1/q})
=K_1$ and suppose that $q \neq p_1$. By considering the prime ideal
factorization of $p_1 p_2$ in $\Q(\zeta_s )$ we may conclude that $(p_1
p_2)^d$ cannot be an $s$th power for any $d=1, \dots,s-1$. Therefore $[K:
\Q(\zeta_s)]=s$.

As we have supposed that $q\nmid c$ there has to be at least one prime $p_3$
that has a ramification index $2$ in the extension $F/\Q$, but is not ramified
in the extension $\Q(\zeta_s)/\Q$. Earlier, we saw that $[K:\Q(\zeta_s)]=s$.
Because $p_3$ is not ramified in $F/\Q$ and $2$ does not divide
$[K:\Q(\zeta_s)]$, none of the prime ideals $P_3$ in $\OO_K$ that lies over
$p_3$ has $2$ as a divisor of the ramification index $e(P_3|p_3)$. This
implies that $F \not\subseteq K$ .

By \cite[Lemma 2]{Perl} we know that $[K_1:K]=q $. Because $q\neq 2$
and $F \not\subseteq K$ the extension $K_1 F /K$ is cyclic and 
$[K_1  F : K]=2q$.

Chebotarev's density theorem \cite[Lemma 7.14, p. 392]{Nark} states that $K$
has infinitely many prime ideals that have absolute degree one and are totally
inert in the extension $K(\sqrt[q]{p_1}) F/K$. We choose one, $P$, that not
only has an absolute degree one but that is also unramified in the extension
$K/\Q$.

We denote the prime of $\Q$ that lies under $P$ by $p$. The field
$\Q(\zeta_{q^n} )$ is a subfield of $K$ and therefore $p$ splits completely in
the extension $\Q(\zeta_{q^n} )/\Q$. The theory of cyclotomic fields \cite[p.
195]{Koch} now gives that
\[p\equiv 1 \pmod{q^n}.\]

Next we are going to show that $p_1^t$ is not an $s$th power $\pmod p$ for
$t=0,\dots, s-1$. We assume the contrary. Suppose that $p_1 \equiv a^q \pmod
p$ for some integer $a$. Now  $p_1 \equiv  a^q \pmod P$. This last equation
however cannot be true because $P$ is totally inert in the Kummer extension
$K_1/ K$. Lemma \ref{index} now states that equation $p_1 \equiv x^t \pmod p$
does not have a solution for any $t=1,\dots, q^n-1$.

Lemma \ref{alkuluku} states that $\Q(\zeta_p)$ has a unique subfield $Z$ with
$[Z:\Q]=q^n$, and that $p_1$ is totally inert in the extension $Z/\Q$.

The prime $P$ has absolute degree one in $K$ and therefore $(p_1 p_2)^{1/q^n}
\equiv c \pmod P$, where $c$ is some integer. This implies that
\[p_1 p_2 \equiv c^{q^n} \pmod p.\]
If we use the notation of Lemma \ref{alkuluku}, the map $\phi$ takes $p_1$ to
the generator $g$ of the group $\Gal(Z/\Q)$ and $p_1 \cdot p_2$ to identity.
The map $\phi$ is a homomorphism and therefore $\phi(p_2)=g^{-1}$, which again
is a generator of the group $\Gal(Z/\Q)$. Lemma \ref{alkuluku} now shows that
$p_2$ is totally inert in the extension $Z/\Q$.

To complete the proof we have to show that the prime $p$ is inert in the
extension $F/\Q$. The prime $P$ must be inert in the extension $F K /K$ and
therefore the prime $p$ has at least some inertia in the extension $FK/\Q$.
Because $p$ is totally split in the extension $K/\Q$ it does not have any
inertia in this extension and therefore Lemma \ref{kompo} states that $p$ must
be inert in the extension $F/\Q$.
\end{proof}

Theorem \ref{main} states that for the center $\Q(i)$ the only problematic
prime power indices are of the form $2^k$. Luckily, the construction of
Corollary \ref{optimiperhe} covers these cases. As a consequence, we can
construct a division algebra with a minimal discriminant for an arbitrary
index. In Table \ref{jakotaulu} we give explicit representations for division
algebras with a prime power index ($<20$) and a minimal discriminant.

For each index $q^n$ we have searched the prime $p$ of the Theorem \ref{main}
along the lines of Example \ref{etsintaalgoritmi}. After the prime $p$ is
found the actual minimal polynomial of the extension $FZ/F\Q$ can be easily
found by considering the subfields of the extension $\Q(\zeta_p)/\Q$. Both
tasks were done with the aid of computer algebra system PARI \cite{PARI2}.

If the center is $\Q(\sqrt{-3})$, the problematic prime powers are $2^n$ and
$3^n$. Algebras of degree $3^n$ we get from Example \ref{otorni}, but degrees
$2^n$ are more problematic. Still for indices $2$ and $4$ we can find suitable
primes even when Theorem \ref{main} is not promising anything. As a conclusion
we can construct a division algebra with a minimal discriminant if the index
is not divisible by $8$.

In Table \ref{tau2} we give explicit representations for our algebras.

\begin{exam}
From Table \ref{jakotaulu} we get that
\[\A_3=(\Q(i)(a_3)/\Q(i) , \sigma_3, (1+i)(2+i))\]
and
\[\A_2=(\Q(i)(a_2)/\Q(i) , \sigma_2, (2+i))\]
are division algebras with minimal discriminants. According to Proposition
\ref{tensoritulo} algebra $\A_2\otimes \A_3 =(\Q(i)(a_6)/\Q(i) , \sigma_2
\sigma_3, (2+i)^5 (1+i)^2)$, where $a_6$ is a zero of the polynomial $x^6 -
2x^5 + (-3i - 51)x^4 + (4i - 30)x^3 + (-2i + 755)x^2 + (-298i + 2134)x + -593i
+ 1628$, is a division algebra of degree $6$ and has a minimal discriminant.
\end{exam}
\vspace*{4pt}

One of the unfortunate properties of our construction is that when we produce
division algebras of a composite index the resulting algebras tend to have
relatively large non-norm elements $\gamma$. In the following example we solve
this problem in one specific case and show that we can always use $\gamma
=(2+i)(1+i)$. The method has a straightforward generalization to more common
situations.
\vspace*{4pt}

\begin{exam}
In what follows we produce the algebra $\A_6$ as a tensor product of two
smaller algebras.

Let $a_2$ be a zero of the polynomial $x^2+i$. The algebra $\B_2=(F({a_2} )/F,
\sigma_2 ,(1+i)(2+i))$ is a slightly modified version of the algebra $\A_2$ of
Table \ref{jakotaulu}. It is a division algebra with a minimal discriminant.

The algebra $\B_3=(F(a_3)/F, \sigma_3, (2+i)^{-1} (1+i)^{-1})$ is a modified
version of the algebra $\A_3$. Proposition \ref{kiransun} gives us that $\B_3$
is still a division algebra. By considering the equation $\B_3\otimes\A_3\sim
M_n(F)$ we see that $\B_3$ has the same discriminant as the algebra $\A_3$.

Because $\B_2$ and $\B_3$ are division algebras with minimal discriminants it
follows from Proposition \ref{tensoritulo} that the tensor product $\A_6=\B_3
\otimes \B_2 =(F(b_2, a_3)/F, \sigma_2 \sigma_3, (2+i)(1+i))$ is a division
algebra with a minimal discriminant. The polynomial $f_6$ is just simply the
minimal polynomial of the generator $a_6$ of the field $ F(b_2, a_3)$.
\end{exam}

\begin{table*}[htp]
\caption{Conductor $p$ of  the cyclotomic field $\Q(\zeta_{p})$, $\gamma$, and
the minimal polynomial $f_n$ of the extension $\Q(i)(a_n)/\Q(i)$}
\label{jakotaulu}
\small
\renewcommand{\arraystretch}{1.44}
\begin{tabular}{|l|r|r|p{13.5cm}|} \hline
$n$ & $p$ & $\gamma$ & $f_n$ \\ \hline
$2$ & & $(2+i)$  & $x^2+i$ \\ \hline
$3$ & $79$ & $(1+i)(2+i)$ & $x^3 + x^2 - 26x + 41$ \\ \hline
$4$ & & $(2+i)$ & $x^4+i$ \\ \hline
$5$ & $11$ & $(1+i)(2+i)$ & $x^5 + x^4 - 4x^3 - 3x^2 + 3x + 1 $ \\ \hline
$7$ & $211$ & $(1+i)(2+i)$& $x^7 + x^6 - 90x^5 + 69x^4 + 1306x^3 + 124x^2 -
5249x - 4663$ \\ \hline
$8$ & & $(2+i)$ & $x^8+i$ \\ \hline
$9$ & $271$ & $(1+i)(2+i)$ & $x^9 + x^8 - 120x^7 - 543x^6 + 858x^5 + 6780x^4 +
7217x^3 - 2818x^2 - 4068x - 261 $ \\ \hline
$11$ & $859$ & $(1+i)(2+i)$ & $x^{11} + x^{10} - 390x^9 - 653x^8 + 52046x^7 +
146438x^6 - 2723930x^5 - 11558015x^4 + 36326009x^3 + 250960565x^2 + 385923388x
+ 145865807 $ \\ \hline
$13$ & $6163$ & $(1+i)(2+i)$ & $ x^{13} + x^{12} - 2844x^{11} - 6017x^{10} +
2908490x^9 + 10238862x^8 - 1340405033x^7 - 6785664624x^6 + 281925130086x^5 +
1909036915713x^4 - 21097272693753x^3 - 192054635052100x^2 - 235667966495418x +
213548387827457$ \\ \hline
$16$ & & $(2+i)$ & $x^{16}+i$ \\ \hline
$17$ & $239$ & $(1+i)(2+i)$ & $ x^{17} + x^{16} - 112x^{15} - 47x^{14} +
3976x^{13} + 4314x^{12} - 64388x^{11} - 136247x^{10} + 422013x^9 + 1631073x^8
+ 411840x^7 - 5840196x^6 - 11894369x^5 - 10635750x^4 - 4739804x^3 - 938485x^2
- 54850x - 619 $\\ \hline
$19$ & $8779$ & $(1+i)(2+i)$& $x^{19} + x^{18} - 4158x^{17} + 8463x^{16} +
6281539x^{15} - 34466097x^{14} - 4291513699x^{13} + 39454551948x^{12} +
1357034568541x^{11} - 17014625218525x^{10} - 184614267432185x^9 +
3035523756071878x^8 + 10088401800577582x^7 - 253111326110358151x^6 -
143208448461319868x^5 + 10612439791376560471x^4 - 3774559232798357892x^3 -
220041647923912963182x^2 + 86083932120501598139x + 1794221202297461499641 
$\\ \hline
\end{tabular}
\end{table*}

\begin{table*}[htp]
\caption{Conductor $p$ of the cyclotomic field $\Q(\zeta_{p})$, $\gamma$, and
the minimal polynomial $f_n$ of the extension
$\Q(\sqrt{-3})(a_n)/\Q(\sqrt{-3})$}
\label{tau2}
\small
\renewcommand{\arraystretch}{1.44}
\begin{tabular}{|l|r|r|p{14cm}|}
\hline
$n$ & $p$ & $\gamma$ & $f_n$ \\ \hline
$2$ & $5$ & $(\sqrt{-3})(2)$ & $x^2 + x - 1$ \\ \hline
$3$ & & $(2)$ & $x^3 - 3x + 1$ \\ \hline
$4$ & $5$ & $(\sqrt{-3})(2)$ & $x^4 + x^3 + x^2 + x + 1$ \\ \hline
$5$ & $101$ & $(\sqrt{-3})(2)$ & $x^5 + x^4 - 40x^3 + 93x^2 - 21x - 17$
\\ \hline
$7$ & $197$ & $(\sqrt{-3})(2)$ & $x^7 + x^6 - 84x^5 - 217x^4 + 1348x^3 +
3988x^2 - 1433x - 1163$ \\ \hline
$8$ & & & \\ \hline
$9$ & & $(2)$ & $x^9 - 9x^7 + 27x^5 - 30x^3 + 9x + 1$  \\ \hline
$11$ & $353$ &  $(\sqrt{-3})(2)$ & $x^{11} + x^{10} - 160x^9 - 525x^8 +
6066x^7 + 26034x^6 - 48369x^5 - 265374x^4 - 42966x^3 + 405001x^2 + 63189x -
170569 $   \\ \hline
$13$ & $4889$ & $(\sqrt{-3})(2)$ & $x^{13} + x^{12} - 2256x^{11} + 15535x^{10}
+ 1555245x^9 - 20301911x^8 - 255557592x^7 + 4688166666x^6 + 3148489502x^5 -
327998691680x^4 + 1203189132463x^3 + 3781862679467x^2 - 26224493395483x +
33207907136809$ \\ \hline
$16$ & & & \\ \hline
$17$ & $9011$ & $(\sqrt{-3})(2)$ & $x^{17} + x^{16} - 4240x^{15}+ 17305x^{14}
+ 5727403x^{13} - 41284287x^{12} - 2705219919x^{11} + 14589308035x^{10} +
564280956214x^9 - 1381250312443x^8 - 51961946136288x^7 + 526852031838x^6 +
1834916754576839x^5 + 1836850197549204x^4 - 23335163152861586x^3 -
34406356236297728x^2 + 60102147038980885x + 73569709231092527$ \\ \hline
$19$ & $8171$ & $(\sqrt{-3})(2)$ & $x^{19} + x^{18} - 3870x^{17} + 41421x^{16}
+ 3724805x^{15} - 43503449x^{14} - 1437461514x^{13} + 12225751511x^{12} +
286728047867x^{11} - 968096767438x^{10} - 28322179217822x^9 -
31203374649750x^8 + 994413740064487x^7 + 3501119135247182x^6 -
8098862899035075x^5 - 59620882192114428x^4 - 90513387045636018x^3 -
3449524754137218x^2 + 73725797301678129x + 35046894150872059$ \\ \hline
\end{tabular}
\end{table*}

\section{An example code and some simulation results}
\label{simulaatiosektio}

One of the ingredients in the construction of the perfect codes was the use of
ideals in improving the shape of the code lattices. In \cite{HL2} we did the
same but for the purpose of saving energy and making the lattice easier to
encode. We include the following simple fact (also known to E. Viterbo,
private communication) explaining why using a principal one-sided (left or
right) ideal instead of the entire order will not change the density of the
code.

\begin{lemma}
Let $\Lambda$ be a maximal order in a cyclic division algebra of index $n$
over an imaginary quadratic number field. Assume that the minimum determinant
of the lattice $\Lambda$ is equal to one. Let $x\in\Lambda$ be any non-zero
element. Let $\rho>0$ be a real parameter chosen such that the minimum
determinant of the lattice $\rho (x\Lambda)$ is also equal to one. Then the
fundamental parallelotopes of these two lattice have the same measure
\[m(\Lambda)=m(\rho (x\Lambda)).\]
\end{lemma}
\vspace*{4pt}

\begin{proof}
By multiplicativity of the norm the minimum determinant of $x\Lambda$ is equal
to the absolute value of $nr(x)$, so the parameter $\rho$ is the unique
positive root of the equation
\[\rho^n | nr(x) | =1.\]
Let us denote this by
\[\rho = \vert \frac{1} {nr(x)} \vert ^{1/n}.\]
On the other hand, the index $[\Lambda : x\Lambda]= | N_{\A/\Q}(x) |$ (see
\cite[Exercise 7, p. 131]{R}). Moreover, \cite[Theorem 9.14, p. 119]{R} tells
us that
\[| N_{\A/\Q}(x)|= |N_{F/\Q}( N_{\A/F} (x))|
\stackrel{\textbf{Remark } \ref{red vs usual}}{=}  |N_{F/\Q}(nr(x)^n)|
 \stackrel{[F:\Q]=2}{=} |nr(x)^n|^2
=|nr(x)|^{2n}.\]
Hence, $[\Lambda:x\Lambda]=|nr(x)|^{2n}$. Scaling the lattice $x\Lambda$ by
the factor $\rho$ will multiply the measure of the fundamental parallelotope
by $\rho^{2n^2}$. The claim immediately follows from these facts by
calculating
\[m(\rho(x\Lambda)) = \rho^{2n^2} m(x\Lambda)
= \vert \frac{ 1}{ nr(x)} \vert ^{2n^2/n}\ [\Lambda : x\Lambda]\ m(\Lambda)
=   \vert \frac{ 1}{ nr(x)} \vert ^{2n}\ | nr(x) |^{2n}\ m(\Lambda)
=m(\Lambda).\]
\end{proof}

We remark that the same fact obviously also holds for principal left ideals of
a maximal order. A way of using the above lemma is that we can choose the
element $x$ in such way that the left (or right) ideal $x\Lambda$ is contained
in the natural order. By moving the code inside the natural order we then to
some extent recover the layered structure of the natural orders, and then,
hopefully, also some of the advantages of the inherent orthogonality between
layers.

For example in the case of the Golden+ algebra we can use the element
$(1-\lambda)^3$ from the ring of integers $\OO_E$ of the larger field
$E=\Q(\sqrt{2+i})$ as a multiplier. Thus, by denoting
\[M=\begin{pmatrix}
(1-\lambda)^3&0\\
0&(1+\lambda)^3
\end{pmatrix}\]
we get the ideal $\I$ consisting of matrices of the form
$aMM_1+bMM_2+cMM_3+dMM_4$, where the coefficients $a,b,c,d$ are Gaussian
integers and the matrices $M_j, j=1,2,3,4$ are from Section \ref{2x2}.
This ideal is a subset of the natural order $\OO_E\oplus u\OO_E$.

\begin{figure}
\includegraphics[width=10cm]{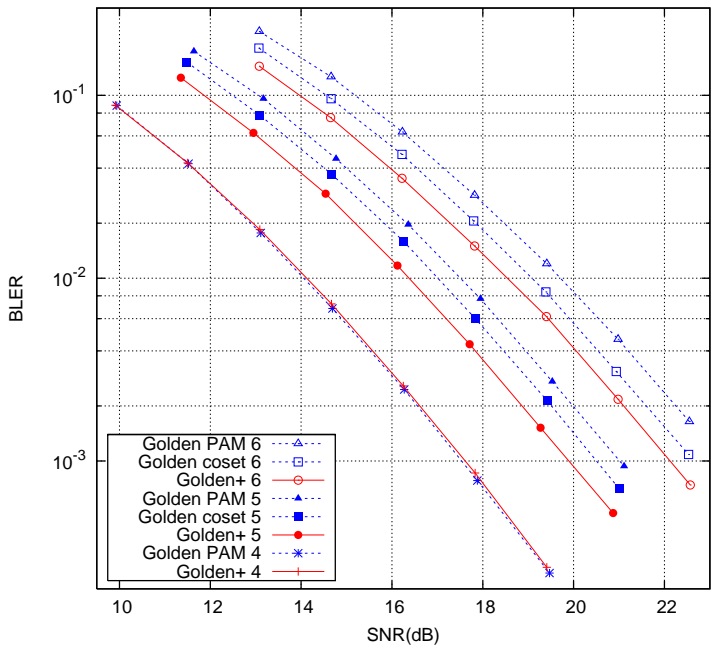}
\caption{Block error rates at 4, 5, and 6 bpcu.}
\label{simulation}
\end{figure}

Our code constructions are based on selecting the prescribed number of lowest
energy matrices from a chosen additive coset of the ideal $\I$. In order to
reach a target bandwidth utilization of $4$, $5$ or $6$ bpcu we thus selected
$256$, $1024$ or $4096$ matrices. In this sense we have done some coset
optimization for the Golden+ codes, but make no claims as to having found the
best coset. For the rival Golden code from \cite{BORV} the coset corresponding
to assigning all the Gaussian integers the value $(1+i)/2$ stands out. This is
because then there are $256$ matrices all having the minimal energy, and more
importantly because in that case  pulse amplitude modulation (PAM) can be used
to good effect. We first did some simulations using a PAM-type rule for larger
subsets of the Golden code as well by arbitrarily selecting a suitable number
of coefficients of the basis matrices from the set $\{-3/2,-1/2,1/2,3/2\}$ so
that the desired bandwidth efficiency was achieved. This is a natural choice
well suited for e.g. the sphere decoding algorithm. While we ended up having a
dead even race BLER-wise at $4.0$ bpcu, the Golden code lost to the Golden+
code by about $0.9$ dB at the higher rates (see Figure \ref{simulation}).
In the interest of a fair comparison we then tried coset optimization for the
Golden code as well. This narrowed down the gap to about $0.3$ dB. However,
the resulting subsets of the Golden code no longer have such a structure well
suited to PAM. In other words both the rival codes must resort to the use of a
code book. We have not even attempted to solve the problem of optimizing the
code book for the purposes of minimizing BER. This also explains, why our
performance plots only show the block error rates (i.e. the probability of
decoder deciding in favor of a $2\times2$ matrix other than the transmitted
one) rather than bit error rates. Thus, our simulations may also be viewed as
measuring the amount of power lost, when one insists on not needing a code
book.

\section{Concluding remarks and suggestions for further work}
\label{conclusion}

We have derived a bound for the density of fully multiplexing MIMO matrix
lattices resulting in codes with a unit minimum determinant. The bound only
applies to codes gotten from the cyclic division algebras and their ideals.
While the bound is not constructive per se, we also showed that it can be
achieved for any number of transmit antennas, and discussed techniques leading
to the construction of CDAs with maximal orders attaining the bound. R.
Vehkalahti is preparing an even more number theoretical article, where these
techniques are expanded. We also discussed the Ivanyos--R\'onyai algorithm
that is needed to actually find these densest possible lattices inside these
CDAs, and gave as an example a construction of a fully multiplexing $2\times2$
code that outperforms the Golden code at least for some data rates.

We have not yet exhausted the box of optimization tools on our code. E.g. the
codes can be pre- and postmultiplied by any complex matrix of determinant one
without affecting neither its density nor its good minimum product distance.
In particular, if we use non-unitary matrix multipliers, the geometry of the
lattice will change. While we cannot turn the lattice into a rectangular one
in this manner, some energy savings and perhaps also shaping gains are
available, but we have not solved the resulting optimization problem yet.
Hopefully a suitably reformed version of our lattice will also allow a
relatively easy description of the low energy matrices. This in turn would
make the use of the sphere decoding algorithm on our lattice more attractive.

There are also possibilities for applying these class field theoretical
techniques to slightly modified density problems of ST-codes. E.g. it is
probably relatively easy to adapt the bound of Theorem \ref{raja} to the case
of multi-block ST-codes. Another possibility is to study the cases, where the
codes are not fully multiplexing. Such situations arise naturally in an
application, where the receiver may have a lower number of antennas, e.g. in a
cellular phone downlink.

An immediate open problem is to utilize maximal orders of the cyclic division
algebra of index $2$ with center $\Q(\omega)$. When looking for the example
code in the previous section a natural step was to use LLL-algorithm for
finding a relatively orthogonal basis for the lattice. That definitely aided
the search for a good coset. In the hexagonal case this step is somewhat
trickier and using a multiplier to put the maximal order inside the natural
order only lead to a code with a disappointing performance. The best way of
using this densest known lattice of $2\times2$-matrices is not known to us. As
another open problem we ask, whether the discriminant bound can be broken by a
MIMO lattice that does not come from a cyclic division algebra. We believe
this to be a very difficult question.

\section{Acknowledgments}
We are grateful to professor Lajos R\'onyai for explaining to us many details
of his algorithm for finding maximal orders.

\end{document}